\documentclass{jfm}
\usepackage{graphicx}
\usepackage{amsmath}
\usepackage{amsfonts}
\usepackage{amssymb}
\usepackage{amsmath}
\usepackage{commath}
\usepackage{epstopdf, epsfig}
\usepackage{siunitx}
\usepackage{hyperref}

\newcommand\Ra{\mbox{\textit{Ra}}} 
\newcommand\Nu{\mbox{\textit{Nu}}} 
\newcommand\Bio{\mbox{\textit{Bi}}} 
\newcommand\Pra{\mbox{\textit{Pr}}} 
\newcommand\Ek{\mbox{\textit{Ek}}} 
\newcommand\Pe{\mbox{\textit{Pe}}} 

\shorttitle{Convection and Robin Boundary Convection}
\shortauthor{T. T. Clarté, N. Schaeffer, S. Labrosse, J. Vidal}

\title{The Effects of Robin Boundary Condition on Thermal Convection in a Rotating Spherical Shell}
\author{Thibaut T. Clarté\aff{1}\aff{2}
    \corresp{\email{thibaut.clarte@univ-grenoble-alpes.fr}},
    Nathanaël Schaeffer\aff{1}
    Stéphane Labrosse\aff{2}
    Jérémie Vidal\aff{1}}
\affiliation{\aff{1} Univ. Grenoble Alpes, Univ. Savoie Mont Blanc, CNRS, IRD, IFSTTAR, ISTerre, 38000 Grenoble, France
\aff{2} Université Claude Bernard Lyon I, ENS de Lyon, LGL-TPE
9 rue du Vercors, Lyon, FRANCE}

\begin{document}

\maketitle

\begin{abstract}
Convection in a spherical shell is widely used to model fluid layers of planets and stars. The choice of thermal boundary conditions in such models is not always straightforward.
To understand the implications of this choice, we report on the effects of the thermal boundary condition on thermal convection, in terms of instability onset, fully developed transport properties and flow structure.
We use the Boussinesq approximation, and impose a Robin boundary condition at the top. This enforces the temperature anomaly and its radial derivative to be linearly coupled with a proportionality factor $\beta$.
Using the height $H$ of the fluid layer, we introduce the non-dimensional Biot number $\Bio^* = \beta H$.
Varying $\Bio^*$ allows us to transition from fixed temperature for $\Bio^* = +\infty$, to fixed thermal flux for $\Bio^* = 0$.
The bottom boundary of the shell is kept isothermal.
We find that the onset of convection is only affected by $\Bio^*$ in the non-rotating case.
Far from onset, considering an effective Rayleigh number and a generalized Nusselt number, we show that the Nusselt and Péclet numbers follow standard universal scaling laws, independent of $\Bio^*$ in all cases considered.
However, the large-scale flow structure keeps the signature of the boundary condition with more vigorous large scales for smaller $\Bio^*$, even though the global heat transfer and kinetic energy are the same.
Finally, for all practical purposes, the Robin condition can be safely replaced by a fixed flux when $\Bio^* \lesssim 0.03$ and by a fixed temperature for $\Bio^* \gtrsim 30$.

\end{abstract}


\section{Introduction}
Thermal convection is the main heat transfer mechanism in the interiors of planets and many stars \citep{phillips2013physics} and is responsible for most of their dynamics. Rayleigh-Bénard convection, which is the flow  developing in a plane layer subject to a potentially destabilising temperature difference, is commonly used as model for the dynamics of such astrophysical objects \citep{Busse76,Busse_inPeltier}. It is also one of the most studied example of pattern--forming system \citep{Cross_Hohenberg93}.

A horizontal layer of fluid heated from below and cooled from above can be unstably stratified -- assuming, as is most usually the case, that the thermal expansion coefficient is positive. Indeed, in that case, the upper cold fluid is denser than the hot one at the bottom. It has been demonstrated \citep{rayleigh1916,chandra} that for a large enough temperature gradient, the fluid adopts a convective state in which heat transport is partially realised by fluid motion.

Convection between two isothermal planes (i.e. with fixed temperature boundary condition, FT) modelled with the Boussinesq approximation constitutes the classical Rayleigh-Bénard thermal convection setup. Lord Rayleigh offered a model for such a phenomenon in 1916 \citeauthor{rayleigh1916} as an explanation for the experiments of \citet{benard1900}.
Other setups of the Rayleigh-Bénard system have been explored, for example by choosing a fixed flux boundary condition (FF), which is closer to usual experimental conditions \citep{busse_riahi_1980}.
However, for geo- and astrophysical applications it can be difficult to choose between those two mathematical usual conditions: fixed field or fixed derivative, also known as Diriclet and Neumann boundary conditions, respectively.
Indeed, in some physical situations, the boundary conditions is more accurately  described by a thermal flux proportional to the temperature itself, leading to
\begin{equation}
    \beta \, T + \frac{\partial T}{\partial r} = 0
    \label{eqbcrob}
\end{equation}
with $\beta$ a proportionality factor homogeneous to an inverse length-scale.
Using a reference length-scale $H$, one can construct the non-dimensional Biot number
\begin{equation}
    \Bio = \beta H.
    \label{eqbiotdef}
\end{equation}
The boundary condition (\ref{eqbcrob}) is intermediate between the Neumann (FF) and Dirichlet (FT) conditions which are obtained for respectively $\Bio=0$ and $\Bio \rightarrow \infty$, and is called the Robin boundary condition.

 Robin boundary conditions appear naturally in the presence of weakly conductive boundaries \citep{busse_riahi_1980} or conductive-radiative empirical Newton's law \citep{o90newton}. Note that linearised black body radiative equilibrium are often interpreted as Newton's law (for example in meteorological models, \citet{savijarvi94}). Thermal contact between two layers of distinct length and thermal conductivity can also be modelled by a Robin boundary condition as done by \citet{guillou95}.

Modelling the radiative equilibrium between a liquid shell and an atmosphere is our initial motivation. More precisely, we are interested in understanding the behaviour of surface global magma oceans crystallizing from the bottom upward \citep{labrosse2015fractional}.
Following the current scientific consensus, Earth may have undergone giant impacts during its formation \citep{togsol}. Such impacts were able to melt a large part of the nowadays solid mantle, producing what is called a magma ocean. 
This physical object is not known with extreme precision but was characterised, at its formation, by large depth ($\sim \SI{1e6}{\metre}$), low viscosity ($\sim \SI{0.1}{\pascal\second}$) and rapid flows ($\sim \SI{10}{\metre\per\second}$) \citep{togsol}.
Modelling the cooling and the crystallization of this terrestrial global magma ocean is of large interest in order to get information about the early evolution of the Earth and other terrestrial planets. In that case, the radiative equilibrium between an atmosphere and the surface of the magma ocean may be modelled using a Robin boundary condition.
Indeed, approximating this radiative equilibrium by an interaction between two black bodies, leads to set the heat flux at the upper boundary of the magma ocean to the difference between the radiative flux leaving the surface and the one received from the atmosphere (assumed isothermal).
Using Stefan-Boltzman's law \citep{boltzman84} (which can be derived from Planck's law) and linearizing the flux variation with temperature leads to the Robin boundary condition.

Up to now, the influence of Robin boundary conditions has received scant attention. Linear stability analyses were conducted in a Cartesian system without rotation by Sparrow et al. \citep{sparrow_goldstein_jonsson_1964}, who determined that increasing the Biot number enables transitioning monotonously from FF to FT cases (in terms of critical Rayleigh numbers). This transition was found to occur in the range $\Bio=$\numrange{0.1}{100}. 

Several studies consider only the extremal cases of FT and FF, the latter being a model for the poorly conductive wall used in some experiments.
In 2D Cartesian geometry, Chapman and Proctor \citeyearpar{chapman80} showed, thanks to stability analysis calculations, the generation of much longer convective cells with symmetric FF, compared to the ones observed in symmetric FT configurations.
A similar 3D setup was also considered by Busse and Riahi \citeyearpar{busse_riahi_1980} who showed that square-pattern convection planforms were preferred to two-dimensional rolls produced in the case of FT boundary conditions.
Later, Ishiwatari et al. \citeyearpar{ishiwatari94} explored, by numerical integration in a Cartesian two-dimensional system, the effects of the different combinations of boundary conditions and internal heating on thermal convection. They observed similar effects on the wavelength of convection cells of the FF boundary conditions (independently of the presence of internal heating) as previously mentioned by Chapman and Proctor.
All these Cartesian geometry studies show that FF produce convection patterns with larger wavelength than FT.

In geophysical fluid dynamics, planetary rotation is often significant and rotational effects (Coriolis force) must be included in the physical model.
In fact, rotation is able to impact dramatically Rayleigh-Bénard convection, altering convection patterns (elongated in Taylor columns) and increasing the intensity of the thermal gradient needed to start convection \citep{chandra}.

Adding the influence of global rotation, Takehiro et al. \citeyearpar{takehiro_2002} studied horizontal layers of fluid with lateral boundaries and rotation axis inclined. They showed that, for a vertical axis of rotation, the behaviour of FF convection at the threshold becomes similar to the FT case when rotation is increased.
In the case of a horizontal rotation axis, the two setups remain different, the FF one being characterised by a zero critical wavenumber.

More recently, Falsaperla et al. \citeyearpar{FALSAPERLA2010122} conducted studies on a Cartesian system with Robin boundary conditions influenced by weak rotation and also observed a monotonuous transition from FF to FT when varying the Biot number.

Influence of Robin boundary conditions has been mainly studied in Cartesian geometry. On the other hand, both rotating and non-rotating convection in spherical shells have been widely studied for its geophysical and planetary relevance but mostly using traditional boundary conditions (FF or FT).

The crystallization of the terrestrial magma ocean is a complex question which cannot be reduced to the influence of Robin boundary condition. Nevertheless, we have chosen to first focus on one of the fundamental aspect of our system of interest. The goal of this work is to study Rayleigh-Bénard convection in a spherical shell subjected to a Robin boundary condition at its upper boundary, with and without rotation.
We first determine numerically the characteristics of the onset of convection with a linear stability analysis, and then we explore the non-linear properties of the same system in terms of convection efficiency.
In all the following we focus first on the two end-member situations, the FF and FT configurations, before studying intermediate cases by varying the value of $\Bio$.
The robustness of these observations is tested by varying the value of the Prandtl number.

\section{Methods}
\label{sec:methods}
\subsection{Physical System}
\label{sec:phymeth}

We consider a spherical shell of aspect ratio $\eta = r_{i} / r_{e} = 2/3$ (all the parameters and fields used are listed in the appendix \ref{sec:anx-nota}), with $r_{e}$ and $r_{i}$ the external and internal radii, respectively, subjected to rotation of vector $\Omega\mathbf{e_z}$ and filled with a Newtonian fluid of constant kinematic viscosity $\nu$, thermal conductivity $k$, reference density $\rho_0$, specific heat capacity $c_{p}$ and thermal expansivity $\alpha$.
The system is subjected to a radial gravity field $\boldsymbol{g}=-g\boldsymbol{e_{r}}$, $g$ being uniform.
Note that our choice of a uniform value is relevant for thin shells and generally speaking for mantle studies (e.g. \citet{bercovici89}), but other radial dependencies of the gravity field can be used in other contexts (e.g. $g \propto r$ for constant density models, or $g \propto \frac{1}{r^{2}}$ for centrally condensed mass, or to conduct exact computations as in \citet{gastine16}).

The physical behaviour of the system is basically monitored by three equations of conservation (conservation of momentum -- the Navier-Stokes equation --, mass and energy) associated with boundary conditions.
We assume the Boussinesq approximation, that is variations of density are neglected except in their contributions to the buoyancy force in which the density varies as function of
\begin{equation}
    \frac{\rho}{\rho_{0}}=1-\alpha(T'-T_{0}),
\end{equation}
$\rho$ being the density, while $\rho_{0}$ and $T_{0}$ are reference values of density and temperature and $T'$ the temperature anomaly around $T_0$. 

The reference state is motionless (uniform zero velocity) and follows a radial conductive temperature profile $T_{ref}$ between the two boundaries with given temperatures (FT)
\begin{equation}
T_{ref}(r) = \Delta T\frac{r_{i}}{r_{e}-r_{i}} \left( \frac{r_{e}}{r}-1 \right) + T_0,
\label{eqthpr}
\end{equation}
where $\Delta T$ is the total temperature difference of the reference profile across the shell.
In the Boussinesq approximation, the value of $T_0$ plays no role and any value can be used such that the reference temperature profile satisfies the Robin boundary condition. In the following, temperatures are defined with respect to this reference.


The problem is rendered dimensionless using $r_{e}$ as length scale, $\frac{\nu}{r_{e}}$ as velocity scale, $\Delta T$ as temperature scale and $\rho_0 \nu \Omega$ as pressure scale.
The system is characterised by the following dimensionless numbers: the Biot number $\Bio$, the Prandtl number $\Pra=\nu \rho_0 c_{p} / k$, the Rayleigh number $\Ra=\frac{g \alpha \Delta T r_{e}^{3}\rho_0 c_p}{\nu k}$ and the Ekman number $\Ek=\frac{\nu}{\Omega r_{e}^{2}}$.

The dimensionless conservation equations are then
\begin{equation}
\frac{\partial \Theta }{\partial t} + \textbf{u}\cdot \boldsymbol \nabla \left( \Theta + \frac{T_{ref}}{\Delta T} \right)=\frac{1}{\Pra} \boldsymbol \nabla ^{2} \Theta
\label{eq_Theta}
\end{equation}
for the energy,
\begin{equation}
\Ek\left(\frac{\partial \textbf{u}}{\partial t}+\textbf{u}\cdot\boldsymbol \nabla \textbf{u} \right) +2 \boldsymbol e_{z} \times \textbf{u} + \boldsymbol \nabla \Pi = \Ek\boldsymbol \nabla ^{2} \textbf{u} + \Ra \Pra \Ek \Theta e_r,
\end{equation}
for the momentum and
\begin{equation}
\boldsymbol \nabla \cdot \textbf{u} = 0,\\
\end{equation}
for the mass, with $\textbf{u},\ \Theta\ \mathrm{and}\ \Pi$ the fluid velocity, the temperature anomaly and the reduced pressure, respectively.

Mechanical boundary conditions imposed on $\textbf{u}$ are non-penetrative ($u_r=0$), stress-free (\textit{i.e.} tangential stress is null, that is with $\ [ \boldsymbol \sigma ]$ the tensor of viscosity stresses of the fluid, $\mathbf{t}$ the tangential and $\mathbf{n}$ normal vector to the interface, $([\boldsymbol \sigma]\cdot\boldsymbol n)\cdot\boldsymbol t=0$ 
at the top ($r=r_e$) and no-slip ($\textbf{u}=\textbf{0}$) at the bottom ($r=r_i$). For the thermal boundary conditions, we impose a fixed temperature at the bottom (FT, i.e. $\Theta(r_{i})=0$) and a Robin boundary condition at the top (i.e. $\Bio \Theta(r_{e}) = \frac{\partial \Theta}{\partial r}(r_{e})$, where $\Bio$ is the Biot number). 

These conditions can be relevant for modelling a fluid undergoing a phase change at its bottom interface and subject to a radiative equilibrium at the top surface. Indeed, pressure at the bottom is set by the fixed depth of the interface, which imposes the value of the temperature thanks to the liquid-solid equilibrium.
Concerning the top interface, radiative equilibrium between a layer of fluid, with surface temperature $T(r_e)$ and surface heat flux $-k \pd{T}{r}(r_e)$, and an isothermal atmosphere at temperature $T_{a}$ leads to an energy balance,
\begin{equation}
\label{eq:rad}
\sigma (T_{a}^{4}-T^{4}(r_e)) = k \pd{T}{r}(r_e),
\end{equation}
where $\sigma \simeq 5.67 \times 10^{-8}\ \mathrm{W.m}^{-2}.\mathrm{K}^{-4}$ is the coefficient of the Stefan-Boltzmann law giving the amount of heat flux produced by a black-body at a given temperature. 

We further introduce the reference depth-dependent temperature $T_{ref}$ such that its value at the surface is set by
\begin{equation}
4\sigma (T_{a}^{4} - T_{ref}^{4}(r_e) ) = k \frac{\partial  T_{ref}}{\partial r}(r_e),
\label{eqto}
\end{equation}
and the depth-dependent temperature anomaly $\theta = T-T_{ref}$. 
Assuming $\theta \ll T_{0}$, equation \eqref{eq:rad} can be expanded to give
\begin{equation}
-4\sigma T_{ref}^{3}(r_e) \theta = k\dpd{\theta}{r}(r_e).
\label{eqbioc}
\end{equation}

Generally speaking, a Robin boundary condition applied on a field $Y$ and its spatial derivative with respect to a space variable $l$ at a boundary $r_0$ can be expressed as $\Bio\, Y+\pd{ Y}{l}=0$. With that definition $\Bio=0$ imposes a FF boundary condition (zero spatial derivative i.e. flux) while $\Bio=+\infty$ imposes a FT boundary condition (temperature anomaly is zero at the interface). 

The Biot number in the usual thermal context is written as $\Bio = h L/k$, with $h$ 
the convective heat transfer coefficient 
having the dimension of a \si{W.m^{-2}.K^{-1}}, $k$ the thermal conductivity of the affected body and $L$ its typical length scale. $\Bio$ can be estimated by different means depending on the type of heat transfer modelled. For material pieces in contact with a conductive-convective atmosphere, the thermal exchanges can be modelled by a thermal Newton's law with $h$ a coefficient of the order of $\SI{10}{W.m^{-2}.K^{-1}}$. As an illustration a meter-size piece of copper in contact with the atmosphere is characterised by $\Bio \simeq 0.1$. Conversely, for a same size polystyrene piece, $\Bio \simeq 1000$. Last example, a 5 cm glass of water cooled by contact with air in usual conditions of pressure and temperature is characterised by $\Bio \simeq 5$. All these estimations can be computed from generic or specific handbooks (e.g. \citet{thhandbook}).

In the case described in eq.~\eqref{eqbioc}, $h=4\sigma T_{a}^{3}$ so that
\begin{equation}
    \Bio = \frac{4\sigma T_{a}^{3}L}{k}.
    \label{defbi}
\end{equation}

Even if our modelling remains unspecific to magma oceans (we basically model a spherical rotating shell of fluid) we should mention the estimated values of the dimensionless numbers characterizing a terrestrial magma ocean modelled by our method. Parameters given are those of a terrestrial fully liquid magma ocean.

Values are computed in both cases and reported in Tab. \ref{dimnumb}.

These parameters constitute extreme conditions for numerical simulations (for example in terms of $\Ek$) which cannot be considered with current computational ressources. Since we focus here on a specific aspect of such a system (modelled by a Robin boundary condition) we restricted our study to a more convenient parameter space (see Tab. \ref{dimnumb}).

\begin{table}
    \centering
    \begin{tabular}{llll}
        Name (symbol) & Magma Ocean & Range in this paper\\

        Ekman number ($\Ek$) & $3 \times 10^{-13}$ to $3 \times 10^{-9}$ & $10^{-4}$ to $\infty$ \\
        Prandtl number ($\Pra$) & $5\times 10^{1}$ to $4 \times 10^{8}$ & $0.03$ to $30$  \\
        Rayleigh number ($\Ra$) & $2 \times 10^{18}$ to $9 \times 10^{25}$ &  $10^{4}$ to $10^{9}$\\
        Biot number ($\Bio$) &$3 \times 10^{5} $  to $3 \times 10^{11}$ &  $0$ to $\infty$\\
    \end{tabular}
    \caption{Dimensionless numbers characterizing a terrestrial global magma ocean, maximal and minimal values are given, computed from \cite{togsol}, \cite{maas2015} and \cite{Snyder94} (for typical values of the thermal conductivity). The extremal values for each parameter cover the full range between a fully liquid to a quasi solid (because of a high proportion of crystals in suspension) magma ocean. Finally, the range of parameters used in this study is precised.}
    \label{dimnumb}
\end{table}

In addition to radiative equilibrium, Robin boundary condition can also model thermal contact between two layers of distinct conductivity ($\lambda_1$ and $\lambda_2$) and thickness ($L_1$ and $L_2$). In that case, following \citet{guillou95}, a Robin boundary condition can be written at the frontier.
It links linearly dimensionless temperature and flux by a factor $Bi = \frac{\lambda_2 L_1 }{L_2 \lambda_1}$.
Such a model could be used for internal oceans of icy satellites of Jupiter \citep{sotin04} or Saturn \citep{collins07} with a layer of liquid water lies below an ice crust.
In that case, the ratio of the conductivities would be typically around 3 (\citet{thhandbook}). This means that a thick ice layer (typically $300$ km) above a thin ocean (typically $10 $km) produces a small $Bi \simeq 0.1$ while a deep ocean below a thin icy crust (reverse thickness ratio) produces a large $Bi \simeq 10$ configuration.

 \subsection{Numerical Tools}
\label{sec:num-tools}
We use the eigenvalue solver SINGE \cite[freely available at \protect\url{https://bitbucket.org/vidalje/singe}; ][]{vidal2015,monville2019rotating} in order to determine the critical Rayleigh number of the system, $\Ra_C$, and, in terms of spherical harmonics, the order of azimutal symmetry $m$ of the most unstable mode at the onset of convection in our system.

SINGE is a Python code conceived to determine the eigenvalues-eigenmodes of incompressible and stratified fluids in spherical cavities. SINGE uses the SLEPc solver \citep{vidal2015}, finite difference ($2^{nd}$ order scheme with $N_{r}$ points) in the radial direction and spherical harmonic decomposition, with degree $l$ and azimuthal wave number $m$ truncated at ($l_{max},m_{max}$). Equatorial symmetry or anti-symmetry is imposed.

\begin{figure}
    \centering
    \begin{tabular}{ll}
        \includegraphics[scale=0.28]{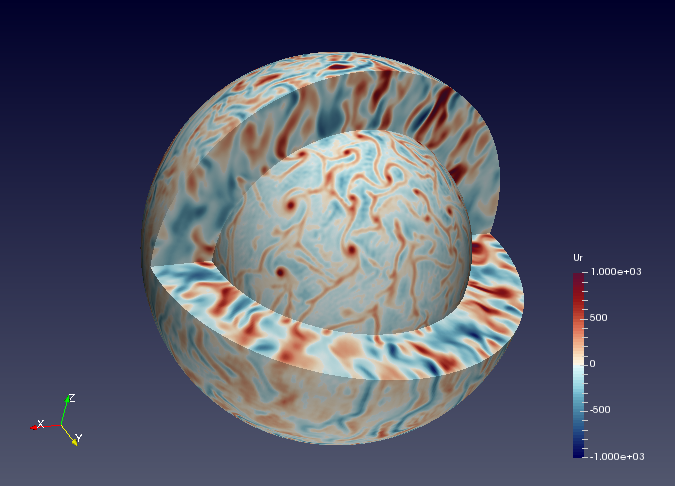} & \includegraphics[scale=0.3]{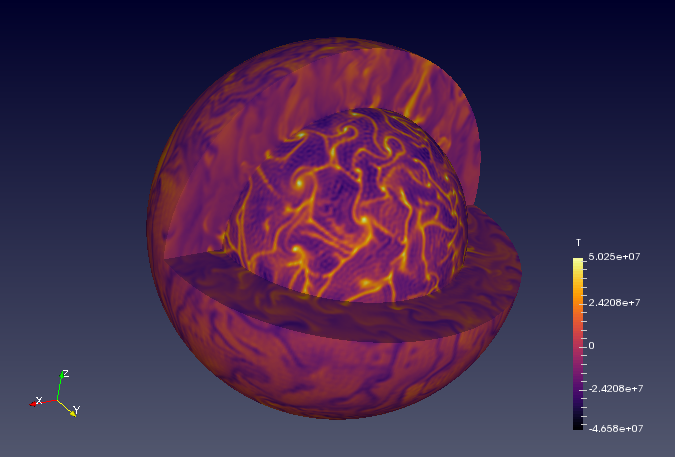}\\
    \end{tabular}
    \caption{3D visualization of radial velocity (left) and temperature anomaly (right) of a nonlinear computation. Simulation run for $\Ra=9\times 10^6$, $\Pra=3$, $\Ek=10^{-4}$ and $\Bio=0$.  Internal and external surfaces have been defined slightly outside boundary layers.}
    \label{3D}
\end{figure}

The non-linear problem is solved by the XSHELLS code (freely available at \url{https://bitbucket.org/nschaeff/xshells}). This geodynamo code has been used to solve convection in spherical rotating shell (see for example \citet{kaplan2017} or \citet{guervilly2019}). XSHELLS solves the Navier-Stokes equation in spherical setup thanks to the toroidal-poloidal decomposition and a pseudo-spectral approach using the fast SHTns library \citep{schaeffer2013}. Poloidal and toroidal scalars as well as temperature are expanded over spherical harmonics truncated in a similar way to SINGE. In the radial direction, XSHELLS uses a second order finite difference differentiation. Each simulation is run long enough to reach a statistically steady state. We then dispose of a 3D-distribution of the anomalies of the main fields (temperature, velocity etc., see fig. \ref{3D} for example). We extract from these latter 2D maps of anomalies or global quantities at a given time or compute time-average data. 

Both numerical codes have been modified for this study in order to take into account the Robin boundary conditions.
 
\section{Linear Stability Analysis}
\label{sec:lin-stab}

\subsection{Differences between FF and FT cases with respect to Ekman number}
\label{sec:amptra}

We determine the critical Rayleigh number and azimuthal wave number (when it is relevant) at the linear onset of convection for different rotation rates ($\Ek=+\infty $ to $10^{-6}$), in both FF and FT end-member cases.
We checked that the symmetric modes are always the preferred unstable modes at the linear onset of convection (i.e.their $Ra_{C}$ is the smallest).

\begin{figure}
\includegraphics[scale=0.7]{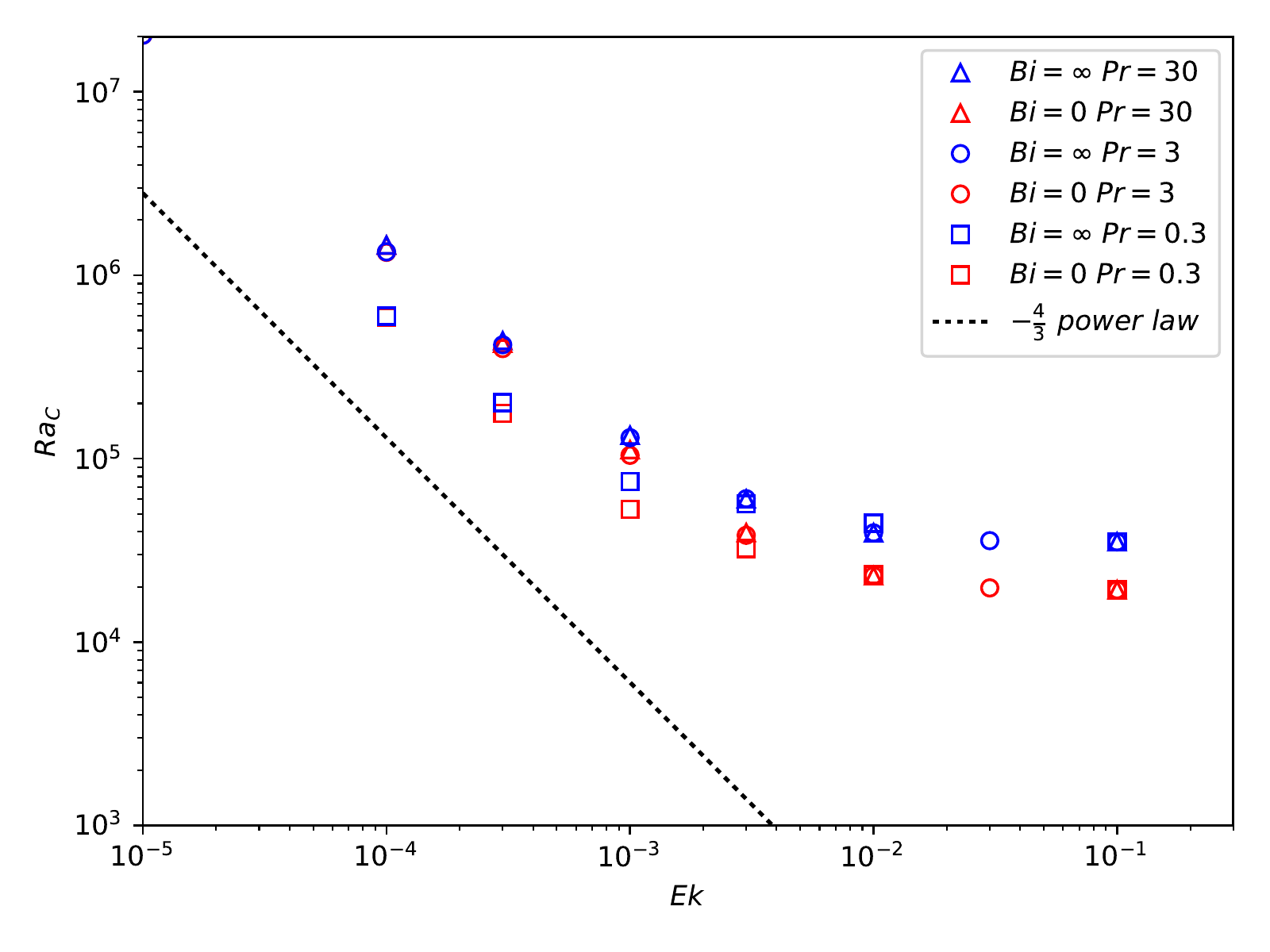}
\caption{Critical Rayleigh number $\Ra_C$ of FF and FT cases as a function of Ekman number $\Ek$ for Prandtl number values $\Pra=0.3$, $3$ and $30$. For evanescent $\Ek$ $\Ra_C$ scales as $\Ek^{-4/3}$ (represented by the dotted line).}
\label{racek}
\end{figure}

In fig. \ref{racek} we plot $\Ra_C$ as a function of $\Ek$, for different values of $\Pra$, in order to illustrate the existence of two regimes (expected from previous studies, e.g. \cite{gastine16}): in the first one ($\Ek \rightarrow +\infty$) rotation does not affect $\Ra_C$ while this number diminishes with $\Ek$ in the second regime. For small enough $\Ek$, the expected power law $\Ra_C \propto \Ek^{-4/3}$ is observed \citep{Chandra1953}. Note that, although $\Ra_C$ is a relevant criterion to discriminate FF and FT setups at high values of the Ekman number, it is no longer the case when rotation impacts notably the critical Rayleigh number. The effect of the Prandtl number is contrasted:  the general behaviour of $\Ra_C$ as a function of $\Ek$ is not affected by the value of $\Pra$ but its influence on the exact value of $\Ra_C$ depends on the Ekman number. For $\Ek \ge 10^{-1}$, $\Pra$ has no visible effect since cases are differentiated only by their boundary conditions. This is in line with the classical behaviour of non-rotating Rayleigh-Bénard convection for which the onset of convection is known to be independent of the value of the Prandtl number \citep[e.g.][]{chandra}. On the other hand, for $\Ek \le 10^{-3}$, the $\Pra=0.3$ cases become distinct from the cases with larger values of $\Pra$ ($\Pra=3$ or $30$).

\begin{figure}
\includegraphics[scale=0.65]{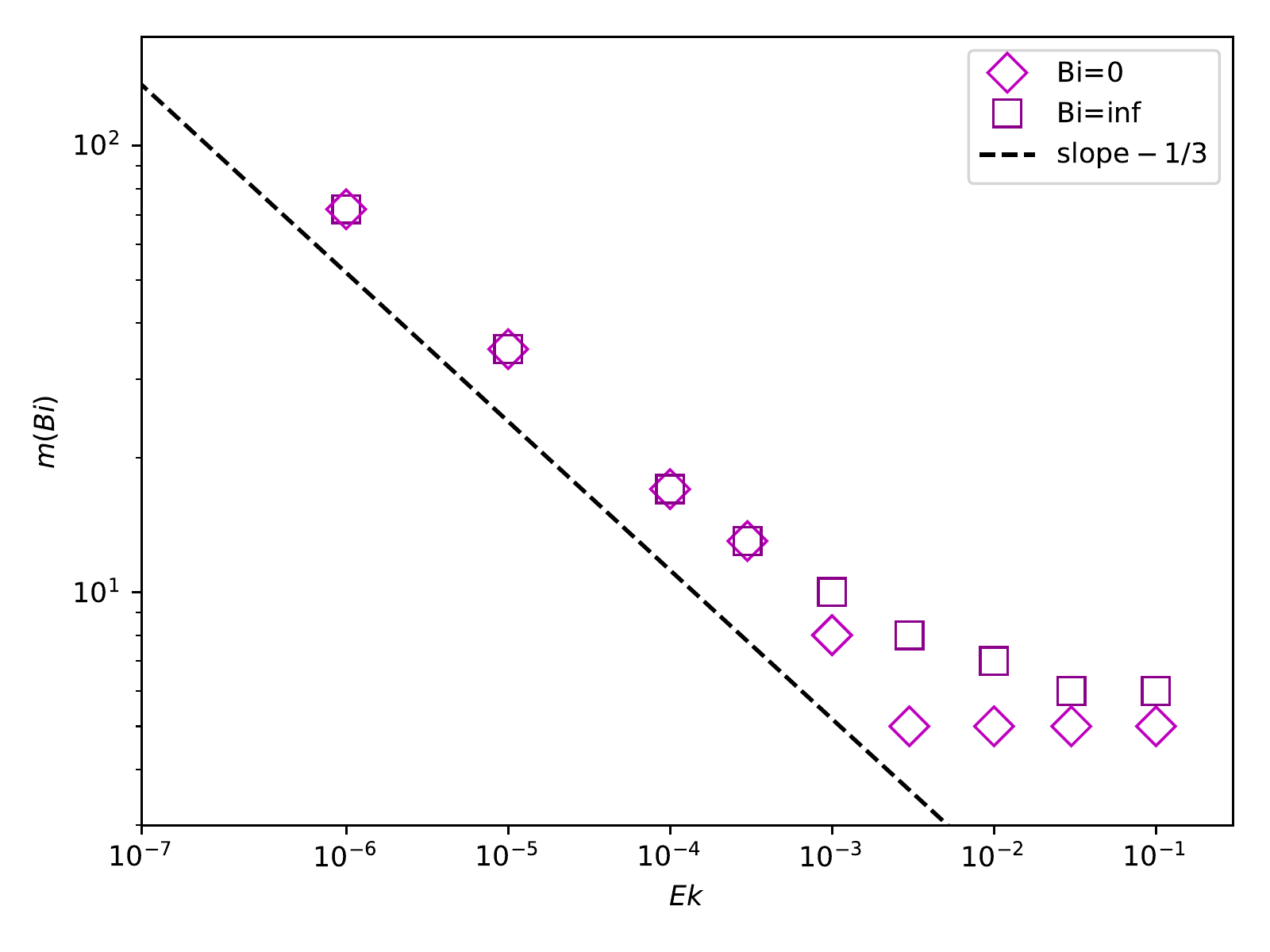}
\caption{Values of the critical azimuthal wave number $m$ as a function of $\Ek$ for both FF ($\Bio=0$) and FT ($\Bio=\infty$) cases. For small enough values of $\Ek$, $m$ scales as $\Ek^{-1/3}$ (represented by the dashed line).}
\label{delm}
\end{figure}

Let us now turn to the azimuthal structure of the flow at the onset of convection. Fig. \ref{delm} shows the critical value of the azimuthal number $m$ for the most unstable mode for both boundary conditions as a function of the Ekman number. We observe no difference between FF and FT configurations for high rotation rates: in the limit of small values of $\Ek$, $m$ scales as $\Ek^{-1/3}$, as expected \citep{busse70}. For large values of $\Ek$, the critical value of $m$ tends toward a constant. In the case of $\Bio=0$, the transition to a constant value of $m$ happens sharply at around $\Ek=\num{2e-3}$. This particularity may be compared to the transition described by Falsaperla et al. \citeyearpar{FALSAPERLA2010122} at $\Ek \simeq 10^{-2}$.

We should also mention that in the non-rotating case, $\Ek=\infty$, 
the $m$ number is not relevant. Indeed in a non rotating setup, the most unstable mode is degenerated in terms of $m$ number. 
Nevertheless, in that case, the critical degree $l$ can be determined and, with our chosen setup, is equal to 5 for a FF boundary condition and 6 for a FT one.

\begin{figure}
\includegraphics[scale=0.7]{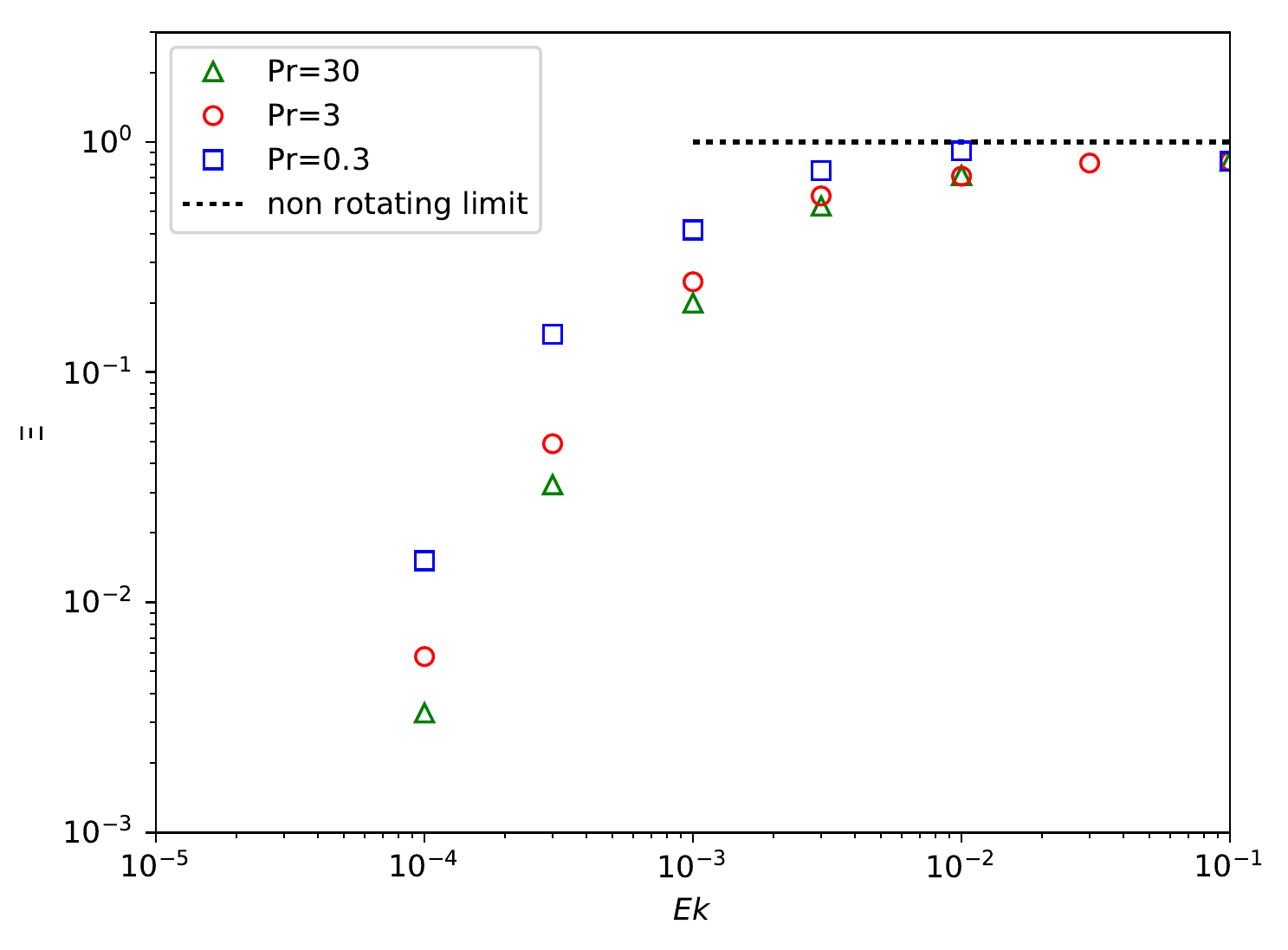}
\caption{$\Xi$ i.e.difference of $\Ra_C$ between FF and FT cases for a given $\Ek$ normalised by the non rotating difference (cf eq. \ref{eqxi}) as a function of $\Ek$. $\Xi$ decreases with the same general behaviour independently from $\Pra$.}
\label{delrac}
\end{figure}

To observe more precisely the distinction between FF and FT cases, we plot in fig. \ref{delrac} 
\begin{equation}
\Xi= \frac{\Ra_C(\Bio=\infty,\Ek)-\Ra_C(\Bio=0,\Ek)}{\Ra_C(\Bio=\infty,\Ek=\infty)-\Ra_C(\Bio=0,\Ek=\infty)},
\label{eqxi}
\end{equation}
which quantifies the normalised difference between the FF and FT cases as a function of $\Ek$. By definition, we have $\Xi \rightarrow 1$ when $\Ek \rightarrow +\infty$.
\label{linpart}

For $\Ek<10^{-3}$, as shown on fig. \ref{delrac}, decreasing the Ekman number decreases the difference between FF and FT cases, independently from $\Pra$. This growing similarity of $\Ra_C(FF)$ and $\Ra_C(FT)$ when rotational effects are increased is in good accordance with the conclusions of \cite{takehiro_2002}. Indeed it has been shown, in the case of top and bottom FF boundary conditions and for Cartesian models in which $\boldsymbol{\Omega}$ is vertical, that when $\Ek$ goes to zero $\Ra_C(FF)\rightarrow \Ra_C(FT)$.

Even though high similarities of $\Ra_C$ and azymuthal periodicity of the flow are observed at high rotation rates between the FF and FT cases, the structure of the solution, in particular its radial shape, depends strongly on the choice of boundary condition. In the FT case, the temperature is uniform at the boundary and lateral heterogeneity is restricted to the bulk of the domain whereas, in the FF case, temperature is variable at the boundary. This is shown later, on figure \ref{dphas}.

After having considered the two extreme regimes (FF: $\Bio \rightarrow 0$ and TF: $\Bio \rightarrow +\infty$) we can now focus on the intermediate case in which the Biot number is arbitrary. 

\subsection{Transition from FF to FT owing to Robin boundary condition}

\label{sec:trlin}

We determine the critical Rayleigh number for different values of the Biot number with various Ekman numbers in the same range as formerly. Increasing $\Bio$ from 0 to $+\infty$ tunes the boundary condition from a fixed flux upper boundary condition to a fixed temperature one (FF to FT transition). We can reasonably expect that the amplitude of this transition is limited by the extremal cases described in section \ref{sec:amptra}. 
Since the convective setup can evolve continuously from FF to FT configuration thanks to the Robin boundary condition, we expect (as the simplest hypothesis) significant influence of this condition on the onset only in the first regime (high Ekman numbers), in which a large $\Ra_C$ difference is observed. 

\begin{figure}
\includegraphics[scale=0.65]{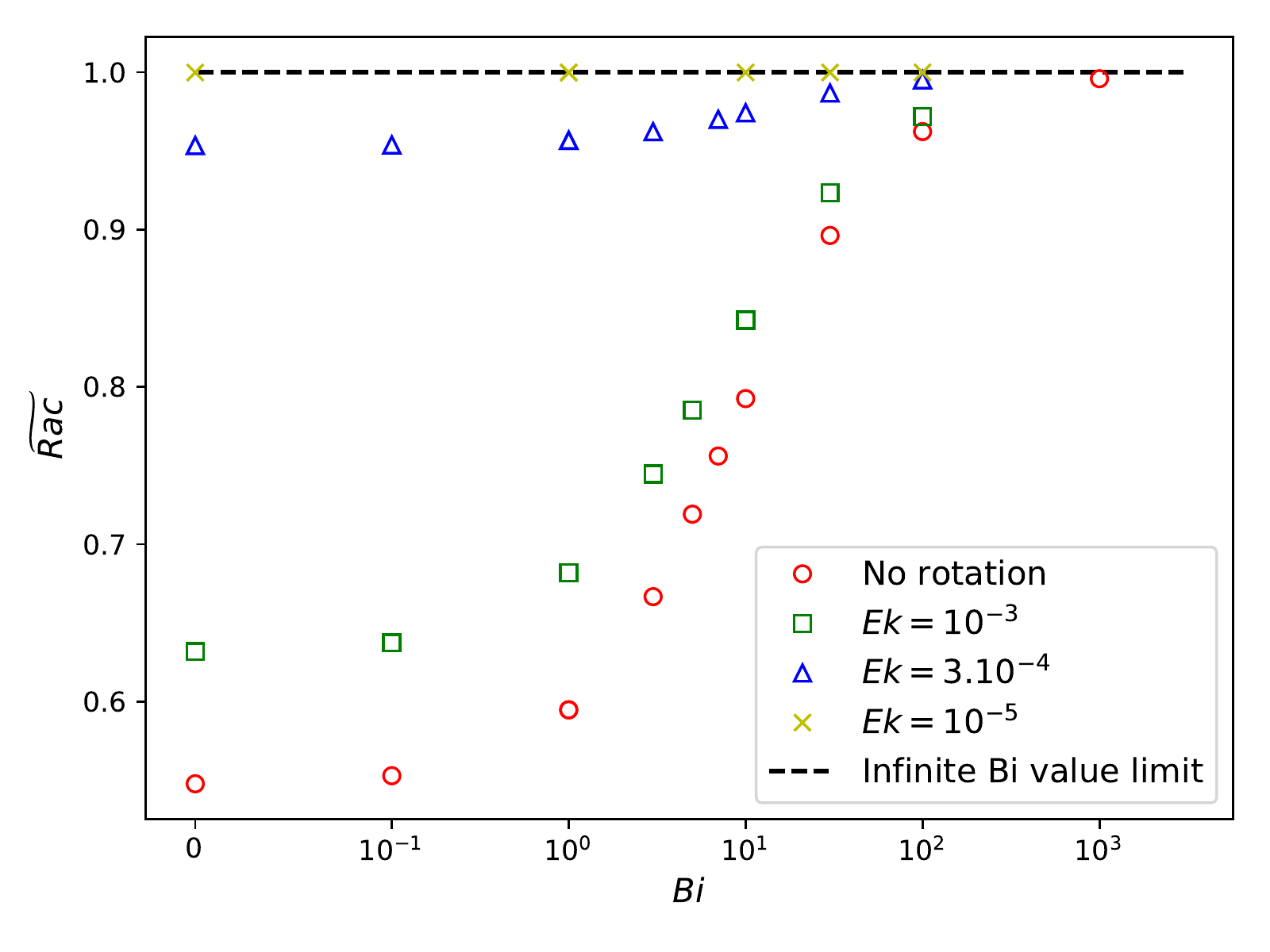}
\caption{ $\widetilde{\Ra_C}$ (cf eq. \ref{ractilde}) as a function of $\Bio$ for $\Pra=3$. The amplitude of the continuous transition from FF to FT is decreased when $\Ek$ is reduced. By definition, $\widetilde{\Ra_C} \rightarrow 1$ when $\Bio \rightarrow \infty$ (i.e.when the boundary condition goes to FT).}
\label{delracbi}
\end{figure}

In order to determine the evolution of the critical Rayleigh number with respect to the Biot number, we plot on fig. \ref{delracbi}, for a given Ekman number, its normalised values,
\begin{equation}
\widetilde{\Ra_C}=\frac{\Ra_C(\Bio)}{\Ra_C(\Bio=\infty)}.
\label{ractilde}
\end{equation}

A continuous and monotonous transition can be observed in terms of the critical Rayleigh number. Without rotation, the effect is the largest, with a minimum ratio of 0.5 ($\Ra_C(FF)=\Ra_C(TF)/2$). As expected from the previous section, the amplitude of the transition is reduced when the Ekman number is decreased. Finally, most of the transition occurs for $\Bio \in [1,100]$.

\begin{figure}
\includegraphics[scale=0.65]{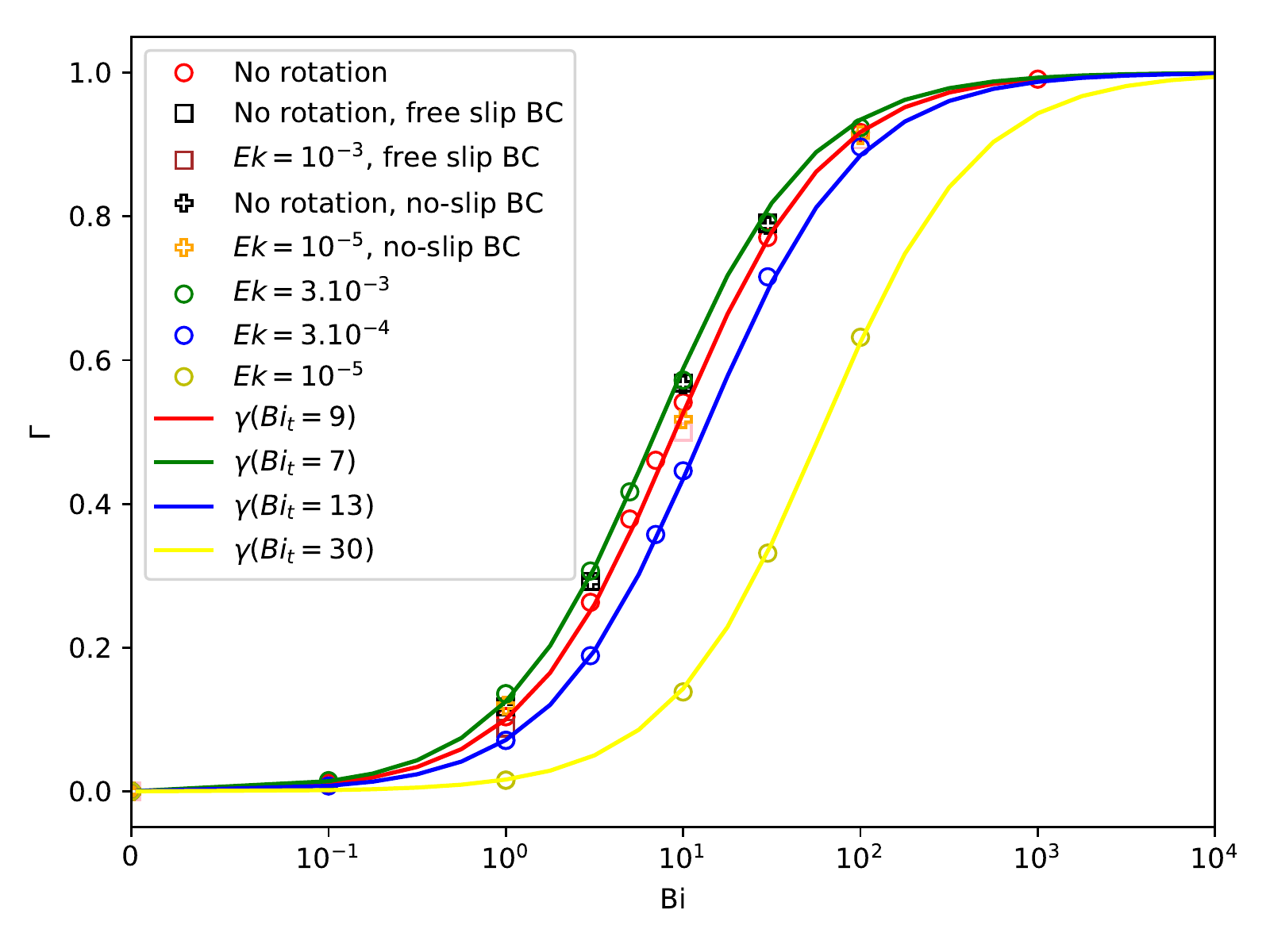}
\caption{$\Gamma = \frac{\Ra_C(\Bio)-\Ra_C(\Bio=0)}{\Ra_C(\Bio=\infty)-\Ra_C(\Bio=0)}$ as a function of $\Bio$ for various velocity boundary conditions at the top of the sphere and values of $\Ek$, $\Pra=3$. It illustrates the transition of $\Ra_C$ from FF values ($\Gamma = 0$) to FT ($\Gamma = 1$) ones. If mechanical boundary condition is mentioned, this one is imposed at both boundaries, if not, the usual setup (free-slip at the top, no-slip at the bottom) is used, see \ref{sec:phymeth}}
\label{delracbin}
\end{figure}

In order to further characterise the transition of the solution as a function of the Biot number, we define
\begin{equation}
\Gamma = \frac{\Ra_C(\Bio)-\Ra_C(\Bio=0)}{\Ra_C(\Bio=\infty)-\Ra_C(\Bio=0)}
\end{equation}
which is plotted as function of $\Bio$ on fig. \ref{delracbin}, for various values of the Ekman number. This quantity allows us to compare the behaviour of the transition regardless of its amplitude set uniformly to 1.
It is worth noting that the transition shows a common behaviour for different combinations of mechanical boundary conditions (symmetric free-slip or no-slip boundary conditions) and rotation rates. The only noticeable difference from one configuration to the other is the location of the transition.
Indeed, we are able to model every transition by
\begin{equation}
\gamma=\frac{\Bio}{\Bio_{t}+\Bio},
 \label{eqmodel}
\end{equation}
where $\Bio_{t}$ is an unknown function of the Ekman number. This model, in which we estimated the suitable $\Bio_{t}$, shows good agreement with data (fig. \ref{delracbin}). We observe that, for moderate rotation rates ($\Ek>10^{-3}$), this threshold is almost constant with $\Bio_{t} \approx 10$. This sets the intermediate regime roughly between $\Bio=1$ and $\Bio=100$. The evolution of the $\Bio_{t}$ parameter with the Ekman number is not monotonous: it first decreases with decreasing $\Ek$ then increases and becomes larger than its value for $\Ek=\infty$ when  $\Ek<10^{-3}$. Nevertheless, as mentioned previously in section \ref{sec:amptra}, the amplitude of the transition --in term of $\Ra_C$-- becomes insignificant with such small values of the Ekman number.

\begin{figure}
\includegraphics[scale=0.7]{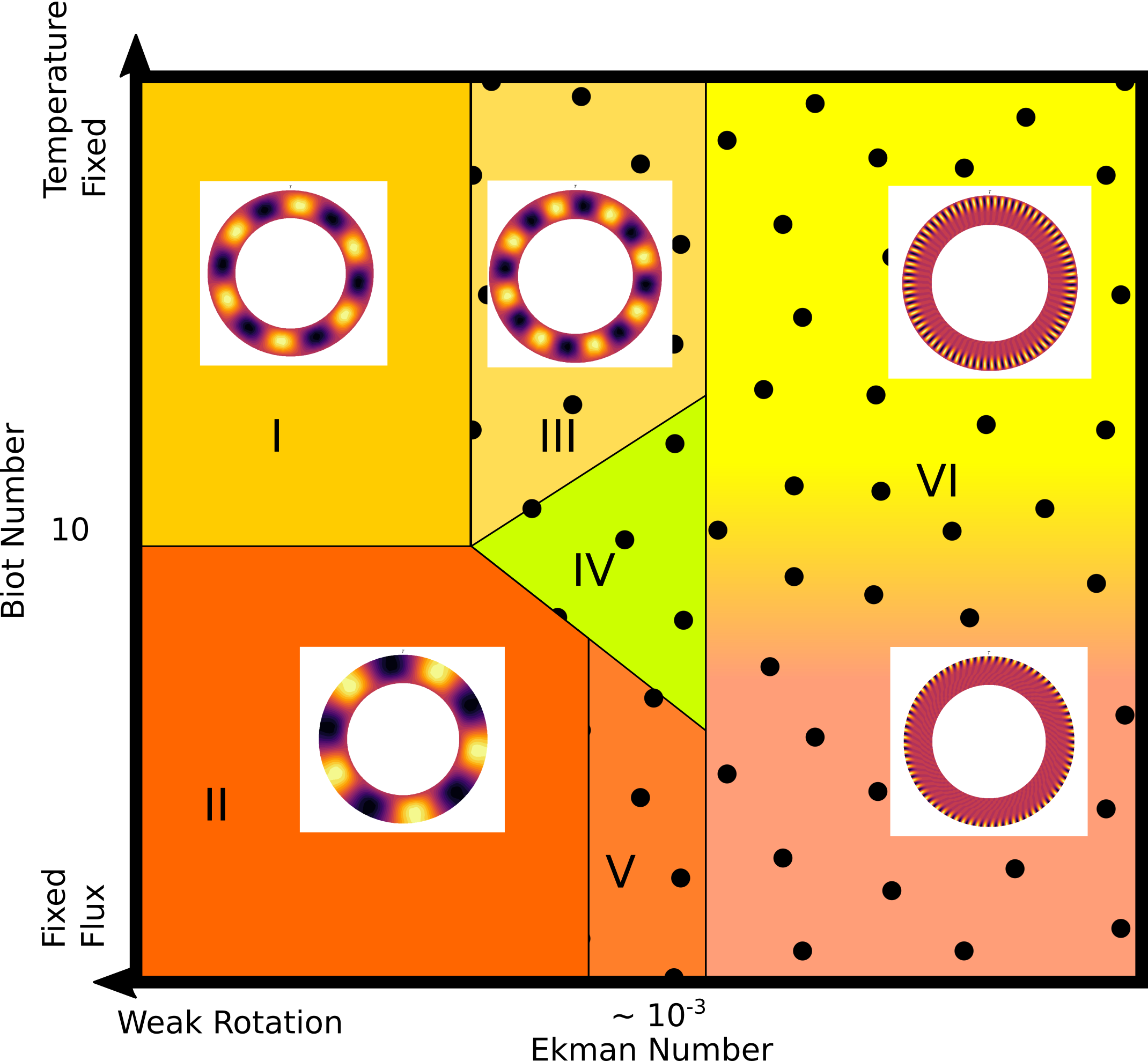}
\caption{Phase diagram of the solution structure at the linear onset of convection as a function of $\Ek$ and $\Bio$, typical structures of temperatures anomalies and dependence on $\Bio$ or $\Ek$ is mentioned for each area. \textbf{I}: FT solution independent of rotation with constant value of $m$, \textbf{II}: FF solution independent of rotation with constant value of $m$, \textbf{III}: FT solution affected by rotation ($m$ increases with rotation), \textbf{IV}: intermediate structure, \textbf{V}: FF solution affected by rotation, \textbf{VI}: common structure at high rotation rate.}
\label{dphas}
\end{figure}

We further discuss the evolution of the solution structure at the onset of convection as a function of the Biot number. Following the results obtained for the two end-member cases of FF and FT situations, discussed in section \ref{sec:amptra}, we first study whether the azimuthal $m$ number of the selected mode depends on the Biot and/or Ekman numbers. This enables us to build a schematic phase diagram of the flow structure at the onset, shown in fig. \ref{dphas}.

In the far left part (high $\Ek$), solutions are weakly affected by rotation and can be described as FF like or FT like structures. When the Ekman number decreases, the solution becomes affected by rotation showing smaller azimuthal number, i.e. a larger $m$ value. Moreover intermediate lateral structures appear for intermediate Biot number, these cases show $m\in [m(\Bio=0),m(\Bio\rightarrow +\infty)]$. Finally, when the Ekman number is very small, the azymuthal symmetry becomes independent of the Biot number. Nevertheless, FF and FT cases remain distinct since the radial extension is still a discriminating diagnostic. Indeed, FF configurations can be characterised by a smaller radial extension of the temperature perturbation compared to FT setups. Also, in the case of FF (and for $\Bio \lesssim 100$), the temperature perturbation extends to the boundary whereas it is zero at the boundary for FT cases.

\subsection{Possible dependence on the Prandtl number}
\label{sec:lintrpr}

We have shown in fig. \ref{delrac}, that the relative difference of $\Ra_C$ measured by the quantity $\Xi$, decreases when $\Ek$ decreases. This general behaviour is not substantially affected by the variation of $\Pra$ in the range $0.3 \leq \Pra \leq 30$ that we investigated.

\begin{figure}
\includegraphics[scale=0.65]{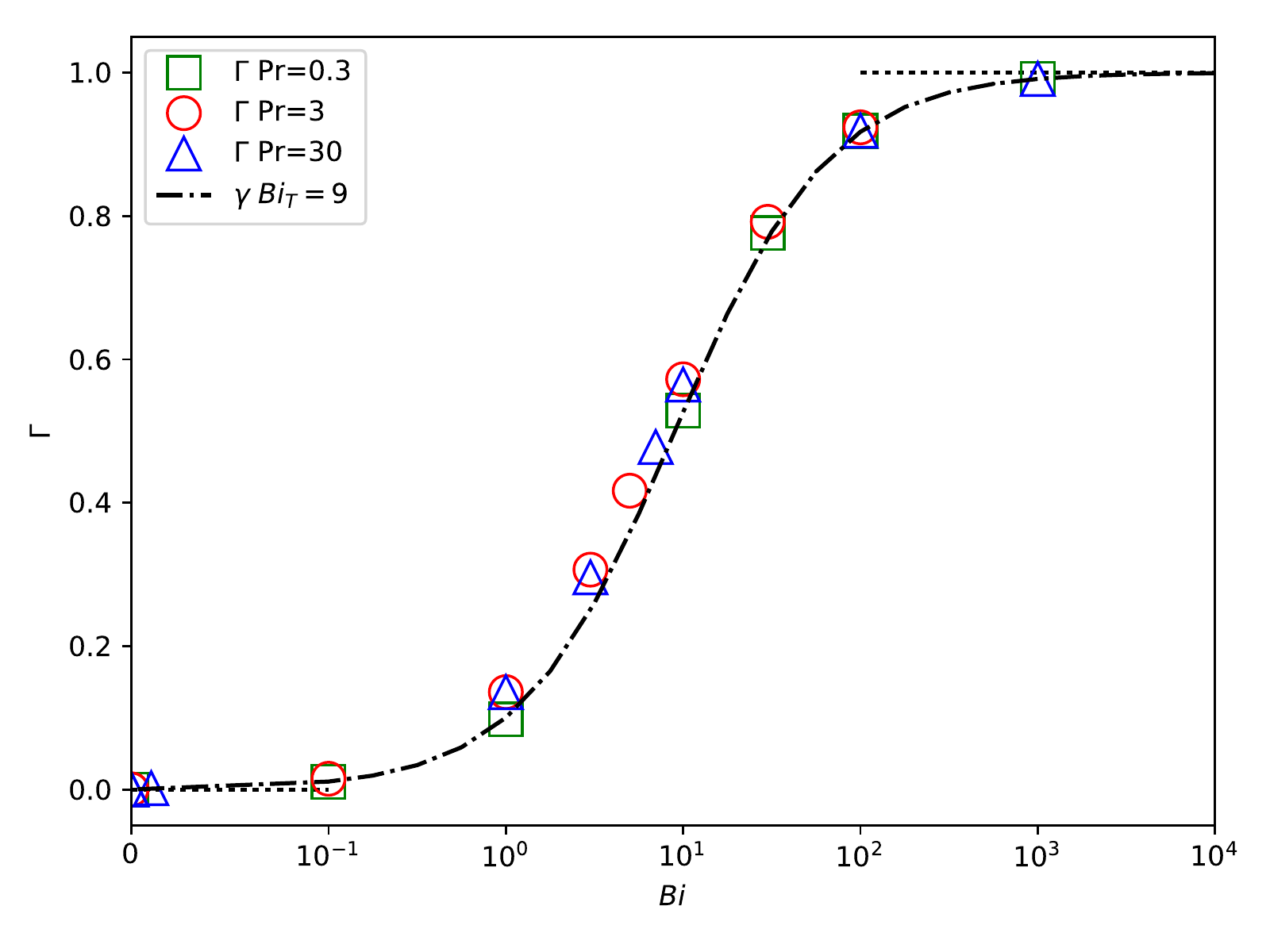}
\caption{$\Gamma = \frac{\Ra_C(\Bio)-\Ra_C(\Bio=0)}{\Ra_C(\Bio=\infty)-\Ra_C(\Bio=0)}$ as a function of $\Bio$ with variable $\Pra$ and $\Ek=+\infty$. $\Gamma$ goes to 0 (respectively 1) when the boundary condition goes to FF, i.e, $\Bio=0$ (respectively FT i.e. $\Bio\rightarrow \infty$).}
\label{trpr}
\end{figure}

In order to discuss the robustness of our observations concerning the effect of $\Bio$ against variations of the Prandtl number, we studied the FF to FT transition in a non-rotating case for $\Pra=0.3$ and $30$. A representation of the transition in terms of critical Rayleigh number is plotted in fig. \ref{trpr}. We can note that the transitional regime is almost independent from $\Pra$. In particular, the range of Biot number over which the transition from FF to FT occurs is  $1 \leq \Bio \leq 100$.

\begin{figure}
    \centering
    \includegraphics[scale=0.3]{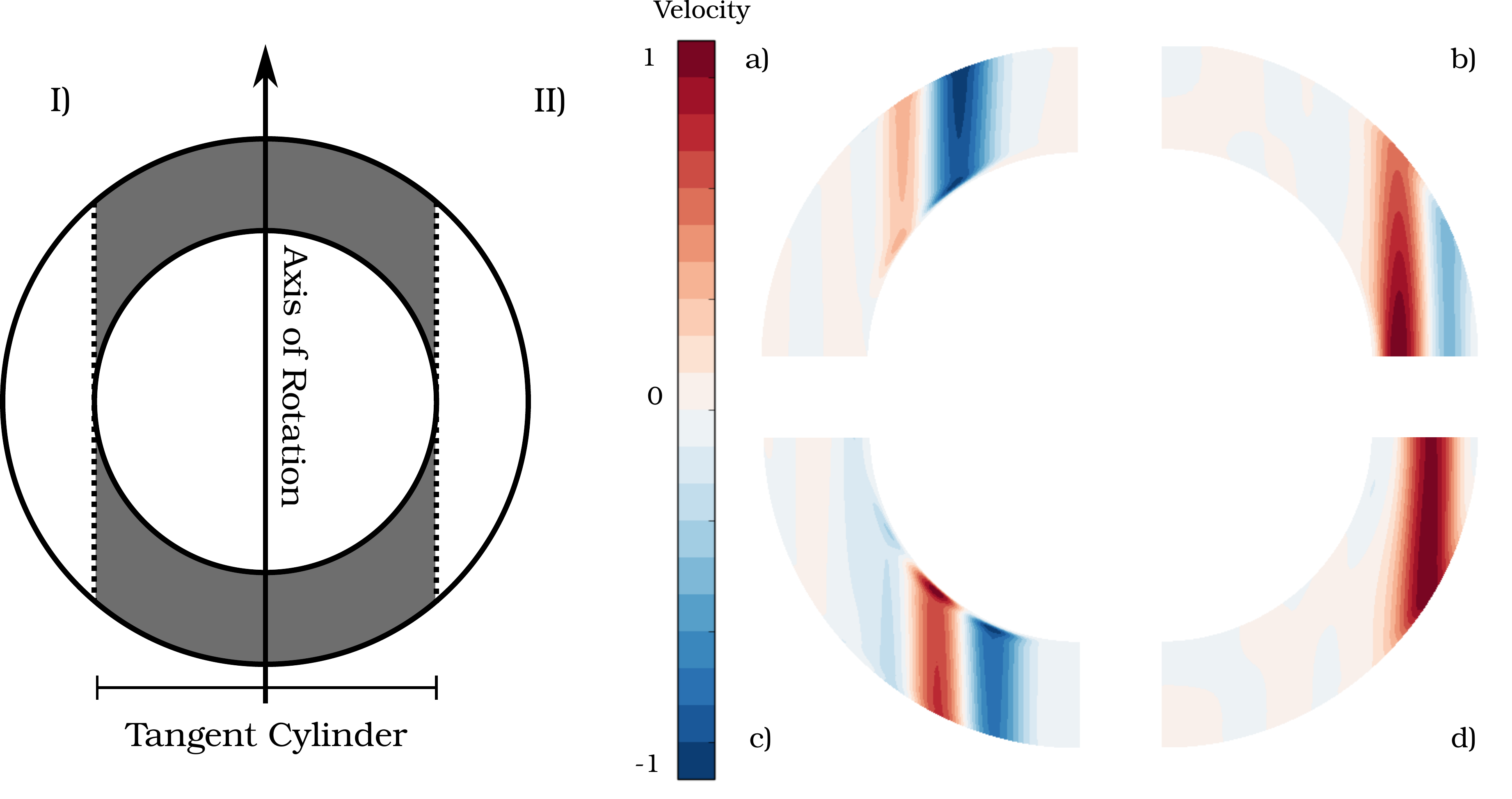}
    \caption{\textbf{I}: Schematic of the tangent cylinder (grey) as a part of a spherical shell rotating about an axis. \textbf{II}: Quarter of meridional slices representing radial cylindrical component of the velocity at the onset of convection for FF (c and d) and FT (a and b) top boundary conditions and $\Pra=0.3$ (b and d) or $0.03$ (a and c), $\Ek=10^{-4}$. We observe an effect of $\Pra$ on the flow, which appears outside of the tangent cylinder for $\Pra=0.3$ and inside it for $\Pra=0.03$. Since these maps are those of the flow at the onset, they are defined within any arbitrary multiplicative factor.}
    \label{fig:U-s-Pr}
\end{figure}

Finally we consider the evolution of the flow structure in a meridional plane when $\Pra$ is varied in the presence of rotation ($\Ek=10^{-4}$).
The radial cylindrical component $u_{s}$ is plotted for FF and FT setups in fig.~\ref{fig:U-s-Pr} II for $\Pra=0.3$ or $0.03$.
Note that for larger value ($\Pra=3$ and 30), patterns similar to the $\Pra=0.3$ case were observed (not shown).
For such large values ($\Pra \geq 0.3$), we observe that the flow at the onset is totally contained outside the tangent cylinder -- the cylinder with axis $\mathbf{\Omega}$ tangent to the inner boundary at the equator, see fig. \ref{fig:U-s-Pr} I.
For lower values of $\Pra$, we observe that the flow at the onset mainly develops within the tangent cylinder. This behaviour seems unaffected by the boundary condition imposed at the top interface.

In other words, when $\Pra$ is decreased, we observe a transition from a symmetric equatorial mode (localised near the equator) to a polar mode (localised near the poles, inside the tangent cylinder).
We can compare our results with those for a thinner shell obtained by \citeauthor{garcia18}. They determined a phase diagram of the selected modes at the onset, as a function of the rotation rate and $\Pra$.
For the same ($\Pra, \Ek$) couples they observed another transition than ours: from an equatorial attached mode to a spiralling columnar one. This difference can possibly be attributed to the different aspect ratio.
Nevertheless in a previous study conducted with a similar shell as ours, \citeauthor{garcia2008} observed a transition from symmetric equatorial mode to anti-symmetric polar one for $\Ek < 10^{-6}$, without mention of polar modes for higher $\Ek$.
Our results suggest that such polar mode can be found up to $\Ek = 10^{-4}$.

\section{Non Linear Simulations}
\label{sec:nl}
In order to investigate the behaviour of the convective flow in the non-linear regime we conducted various non-linear numerical simulations for different values of $\Ra,\, \Ek,\, \Pra$ and $\Bio$ (see section \ref{sec:phymeth}).
We aim to determine if thermal convection subjected to a Robin boundary condition shows specific behaviours for global parameters (convection efficiency or kinetic energy) as well as local ones (structure of the flow). We first consider end-member configurations (FF and FT), before comparing them with solutions for intermediate Biot number.

In the following, we use a global diagnostic, the Nusselt number, which we define as
\begin{equation}
 \Nu=\frac{1-\eta}{\eta} \frac{\widetilde{Q}}{\widetilde{\Delta T_{eff}}}=\frac{1-\eta}{\eta}\frac{Qr_{e}}{\Delta T_{eff} k}   
\end{equation}
where $Q$ is the total heat flux leaving the system and $\Delta T_{eff}$ is the observed temperature difference across the system. The tilde symbolises the dimensionless character of the physical quantities. 
This definition of the Nusselt number is valid in the whole range of boundary conditions (from FF to FT) as imposed by the value of $\Bio$. In order to facilitate the understanding, we note $Q_{ref}$ the heat flux, computed in the motionless reference state (see eq. \ref{eqthpr}), leaving the system at the upper interface.
When $\Bio \rightarrow 0$, in a FF situation, $Q = Q_{ref}$, $\frac{Q (1-\eta)}{r_{e} \eta}=\Delta T $ and $\Nu = \frac{\Delta T}{T(r_{i})-T(r_{e})}$. When $\Bio \rightarrow \infty$, in a FT situation, $T(r_{i})-T(r_{e})=\Delta T$, $ \frac{\eta \Delta T}{r_{e} (1-\eta)}= Q_{ref}  $ and $\Nu = \frac{Q}{Q_{ref}}$. In both end-member cases, the usual definitions of $\Nu$ for FF (see for example \citet{long2020}) or FT (see for example \citet{plumley2019}) setups are obtained and $\Nu$ provides a dimensionless measure of the efficiency of heat transfer in all cases.
According to this definition, $\Nu = 1$ in the absence of convection. Consequently $\Nu-1$ will be preferred on some plots where $\Ra \approx \Ra_C$. $\Nu$ is used to quantify the efficiency of convection in the non-linear system in the statistically stationary regime, consequently its average over time will be presented in this regime.

\subsection{Behaviour of purely FT or FF setups}
\label{sec:extrnl}
\subsubsection{Global diagnostics}

\label{sec:nlgdiag}

First of all, we study end-member configurations in terms of $\Bio$ values, that is $\Bio=0$ or $\Bio\rightarrow + \infty$, for rotating ($\Ek=10^{-4}$) or non-rotating cases, for $\Pra=0.3$ or $3$. The Rayleigh number has been set between $\Ra \approx \Ra_C$ and $\Ra / \Ra_C \approx 10^{3} $. As expected from numerous previous studies \citep[e.g.][]{gastine16,long2020}, several successive convection regimes are observable when $\Ra$ increases above $\Ra_C$.

\begin{figure}

\includegraphics[scale=0.65]{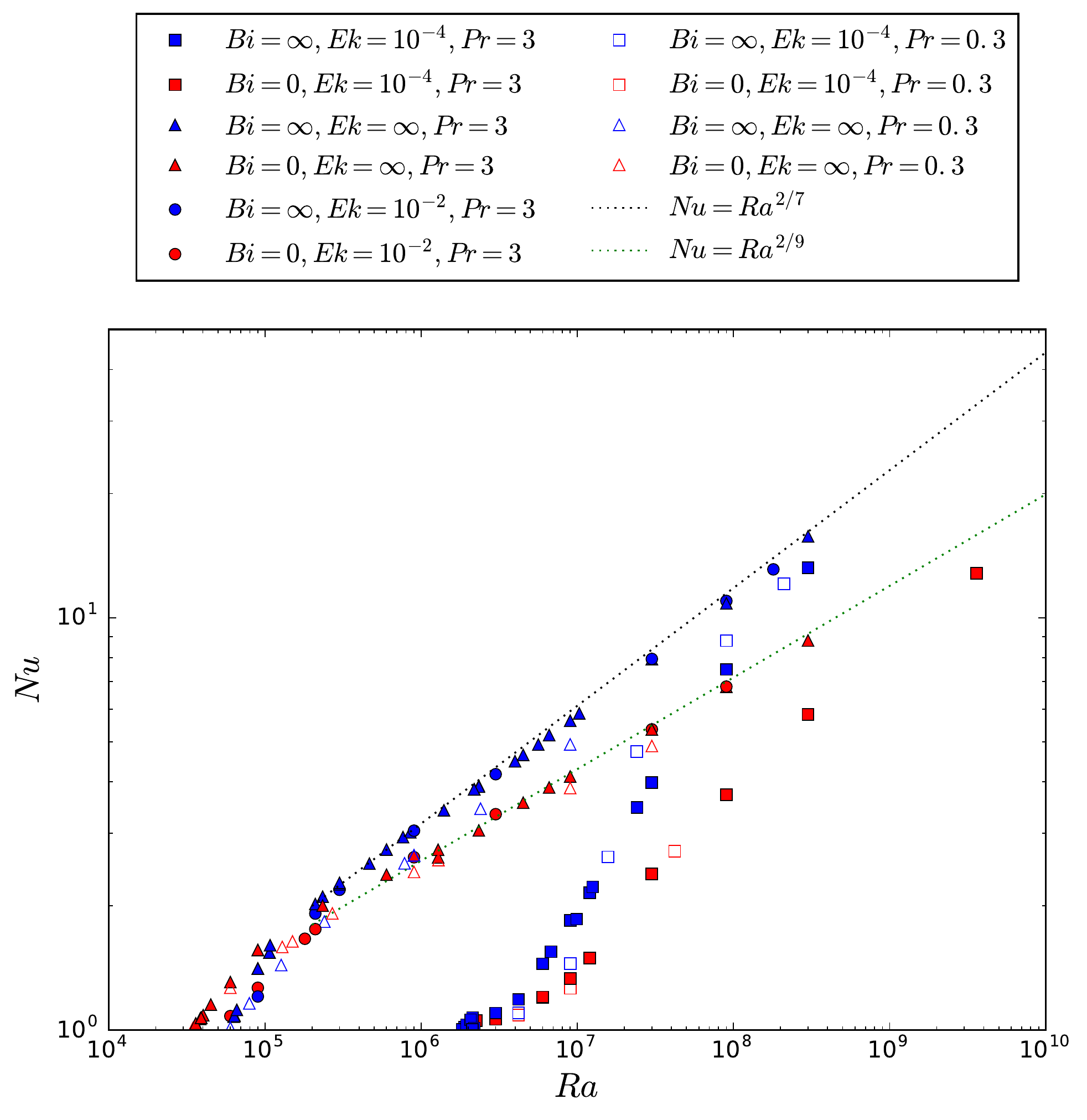}
\caption{Nusselt number, $\Nu$, as a function of $\Ra$ for $\Pra=0.3$ (open symbols) or $3$ (filled symbols) with ($\Ek=10^{-4}$: squares and $\Ek=10^{-2}$: circles) or without ($\Ek=\infty$ circles) rotation and with fixed flux (red points) or fixed temperature (blue ones) at the upper boundary. The red dotted line represents the scaling $\Nu \propto \Ra^{2/7}$ relevant to model FT configurations while the $\Nu \propto \Ra^{2/9}$ --blue dotted line-- models the FF ones.}
\label{nu}
\end{figure}

We first consider $\Nu=f(\Ra)$ as represented on fig. \ref{nu}. We observe the well known convective behaviour (e.g. \cite{King2012} or \cite{gastine16}): an increase of $\Nu$ with $\Ra$ starting from $\Nu=1$ for $\Ra\le\Ra_C$. Rotating cases show a higher $\Ra_C$ and a smaller $\Nu$ at a given $\Ra$ compared to the non-rotating cases. Since the slope of $\Nu=f(\Ra,\Ek \neq \infty)$ is also steeper (it can approach 1/2 for $\Ra \approx 10^{7}$) rotating and non-rotating cases show increasingly similar $\Nu$ when $\Ra$ is increased. We do not reach large enough values of $\Ra$ to see the convergence of moderately rotating and non-rotating cases that has been documented \citep{King2012,Gastine2015,gastine16}. Indeed the convective Rossby number, $R_{oc}=(\frac{\Ra \Ek^{2}}{\Pra})^{1/2}$, used to compare \textit{a priori} the Coriolis force to the buoyancy force \citet{gilman77}) of $\Ek=10^{-4}$ cases is always below 10 (see appendix \ref{sec:anx-zf}). Nevertheless, for weakly rotating cases, this convergence is well observed for $\Ra > 10^6$. 
The comparison of FT and FF cases on the basis of $\Nu$ values at given $\Ra$ shows that the two configurations almost always show distinct behaviour. Generally speaking $\Nu(FF)<\Nu(FT)$. The only exception is noted for the $\Ek=10^{-4}$ cases when $\Ra \approx \Ra_C$ which shows no real difference between FF and FT configurations. This agrees with the fact that the $\Ra_C$ of these cases were found with the linear stability analysis to be very similar. Indeed, it is expected \citep{busse1986,gillet2006} that, near the threshold, the behaviour of the system is controlled by $\frac{\Ra}{\Ra_C}-1$.
In the non-rotating case, for large enough $\Ra$, it is known that the Nusselt number scales as $\Ra^{\alpha}$ where $\alpha$ is expected to be similar to 1/3 (1/3 in \citet{plumley2019} for example or 2/7 are both frequently used, see for example \citet{Gastine2015} or \citet{Iyer2020} for FT configuration and \citet{long2020}, \citet{King2012} or \citet{Johnston2009} for symmetric FF one). Here, we observe $\alpha = 2/7$. This behaviour is well observed in our data set for the FT configuration. Nevertheless, a different exponent ($2/9$ here, as justified below) seems to be necessary to model the FF data. 
All these general observations can be applied to both $\Pra=0.3$ and $\Pra=3$ cases.

The differences observed in fig. \ref{nu} between FF and FT cases ($\Nu(FT)>\Nu(FF)$ at a given $(\Ra,\Pra,\Ek$)) can be explained by fundamental differences of the setups. In the first case, the stationary state is characterised by an effective difference of temperature $\Delta T_{eff}$ smaller than the fixed $\Delta T$ imposed in the FT case. To compare the values of $\Nu$ produced by FF and FT cases on an equal footing, we measure the effective Rayleigh number $\Ra_{eff}$ established in the stationary state defined as (see \citet{calkins2015,Johnston2009,verzicco2008}):
\begin{equation}
    \Ra_{eff}=\frac{\Ra\, \Delta T_{eff}}{\Delta T}.
    \label{eqraeff}
\end{equation}
One can note that $\Ra_{eff}=\Ra$ when FT boundary condition is imposed at the upper interface.
In order to precisely compare FF and FT configurations at the same $\Ra_{eff}$ we first run a FF case then determine its $\Ra_{eff}$ in order to run a FT case at the same $\Ra=\Ra_{eff}$.

\begin{figure}
\includegraphics[scale=0.65]{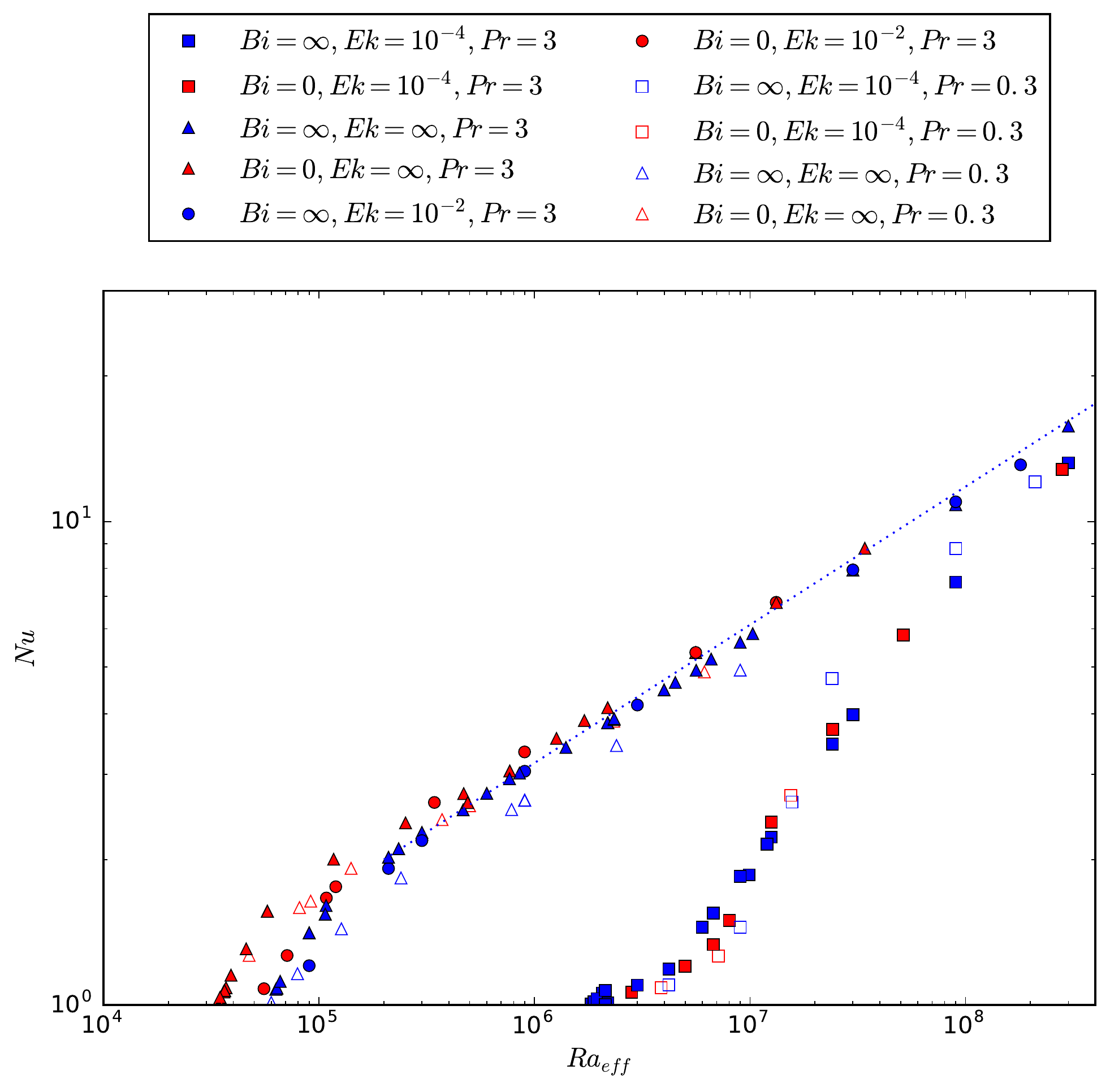}
\caption{$\Nu$ as a function of $\Ra_{eff}$ (\ref{eqraeff}) for the same data as in fig. \ref{nu}. For a given $\Pra$, FF and FT points follow the same behaviour except near the threshold for non-rotating convection. $\Nu\propto \Ra^{2/7}$ is plotted with a blue dotted line.}
\label{Ra_eff}
\end{figure}

We plot, in fig. \ref{Ra_eff}, $\Nu$ as a function of $\Ra_{eff}$. One can notice that, for a given ($\Pra, \Ek$), $\Nu(\Ra_{eff})$ follows the same behaviour independently of the top boundary condition except for cases near the threshold for non-rotating convection. This is reminiscent of the behaviour observed with the linear stability approach (\S\ref{sec:lin-stab}) that showed that the threshold is highly dependent on the boundary conditions only in the absence of rotation. As expected, a fixed-flux boundary condition implies a lower critical Rayleigh-number (which has similar implication on the critical $\Ra_{eff}$ since $\Ra \simeq \Ra_{eff}$ for $\Ra \simeq \Ra_C$). Finally, as shown in previous studies \citep[e.g.][]{calkins2015} FF and FT cases become indistinguishable for high enough values of $\Ra_{eff}$ (here typically $\Ra_{eff} > 3\times 10^5$), in terms of $\Nu$ and are in good agreement with the $2/7$ power law mentioned above.  

\label{compglob}

Since one can write, in the FF case, $\Ra_{eff,FF} \propto \frac{\Ra}{\Nu_{FF}}$, the new scaling $\Nu_{FF} \propto \Ra_{eff}^{2/7}$ leads to $\Ra_{eff,FF}\propto \Ra_{FF}^{7/9}$ and $\Nu_{FF}\propto \Ra^{2/9}$ which was observed on fig. \ref{nu}. Using our data to determine the coefficient of proportionality between $Ra_{eff,FF}$ and $Ra_{FF}$, we found for non rotating convection and $\Pra=3$ that, far enough from the onset of convection and along at least three orders of magnitude,
\begin{equation}
    Ra_{eff,FF} \simeq 8.6 \times Ra_{FF}^{7/9},
    \label{ffraeff}
\end{equation}
with a discrepancy below $5 \%$.

\begin{figure}
\includegraphics[scale=0.65]{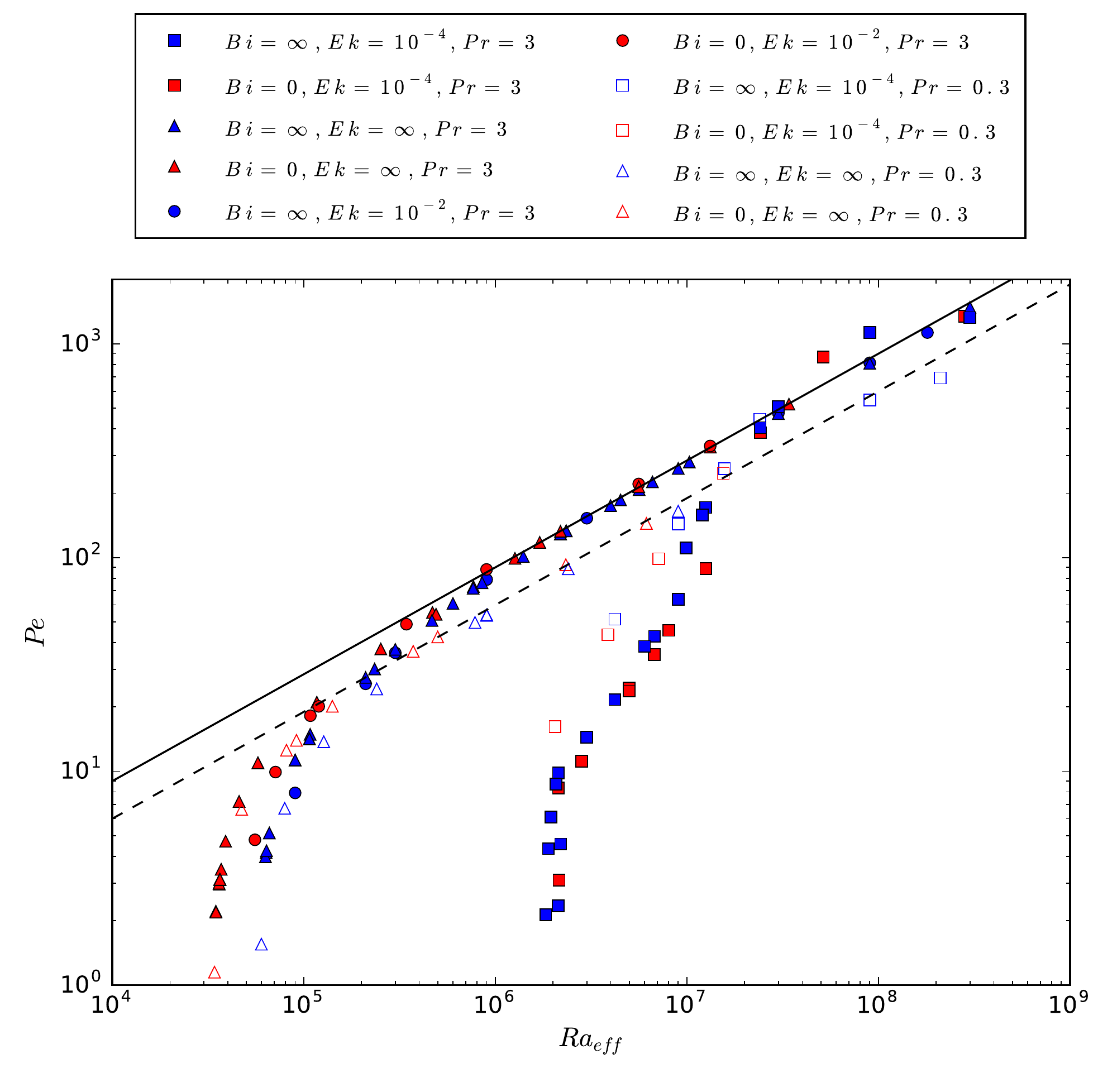}
\caption{Péclet number $\Pe$ as a function of $\Ra_{eff}$ for the same data as plotted in fig. \protect\ref{Ra_eff}. Except for $\Ra_{eff} \simeq \Ra_C$, FF and FT configurations follow the same behaviour at a given $\Ek, \Pra$. For large enough $\Ra_{eff}$, $\Pe$ becomes independent of $\Ek$. Solid and broken lines are $1/2$ slope scalings for $\Pra=3$ and $\Pra=0.3$.}
\label{peRa_eff}
\end{figure}

The similarity of the FF and FT configurations observed at equal $\Ra_{eff}$ on the basis of $\Nu$ prompted us to look at other quantities. Aside from determining $\Nu$ of each configuration, we computed their Péclet Number in statistically steady state, $\Pe=\frac{r_{e}\, u \rho_0 c_p}{k}$, where $u$ is the rms flow velocity.
As one can see on fig. \ref{peRa_eff}, where $\Pe$ is plotted as a function of $\Ra_{eff}$, for a given $(\Ek,\Pra)$ the general behaviour far from the onset of convection is independent from the top boundary condition. For the rotating case, $\Pe$ increases sharply with $\Ra_{eff}$ near the onset, making it difficult to observe low values of $\Pe$.
$\Pra=0.3$ cases can be distinguished by a systematically smaller value of $\Pe$ for $\Ek=\infty$ while they follow the same general behaviour as for the $\Pra=3$ points for rotating cases. We also observe the decrease of the gap between rotating and non-rotating configurations for high values of $\Ra_{eff}$. They almost reach a unified regime characterised by a $\Pe\propto \Ra^{1/2}$ scaling as expected from \citet{Gastine2015} or \citet{long2020} (since $\Pe$ can be deduced from the Reynolds number by the multiplication by a factor $\Pra$ independent of $\Ra$). We plot two models $\Pe\propto \Ra^{1/2}$, one for each $\Pra$ value. Note that the difference of the proportionality factor between the models is in good accordance with a $\Pe\propto \Pra^{1/2}$ dependency --that is a $\Rey \propto \Pra^{-1/2}$ relation found with a somehow different setup in \citet{calzavarini2005}, for example.

\begin{figure}
\includegraphics[scale=0.65]{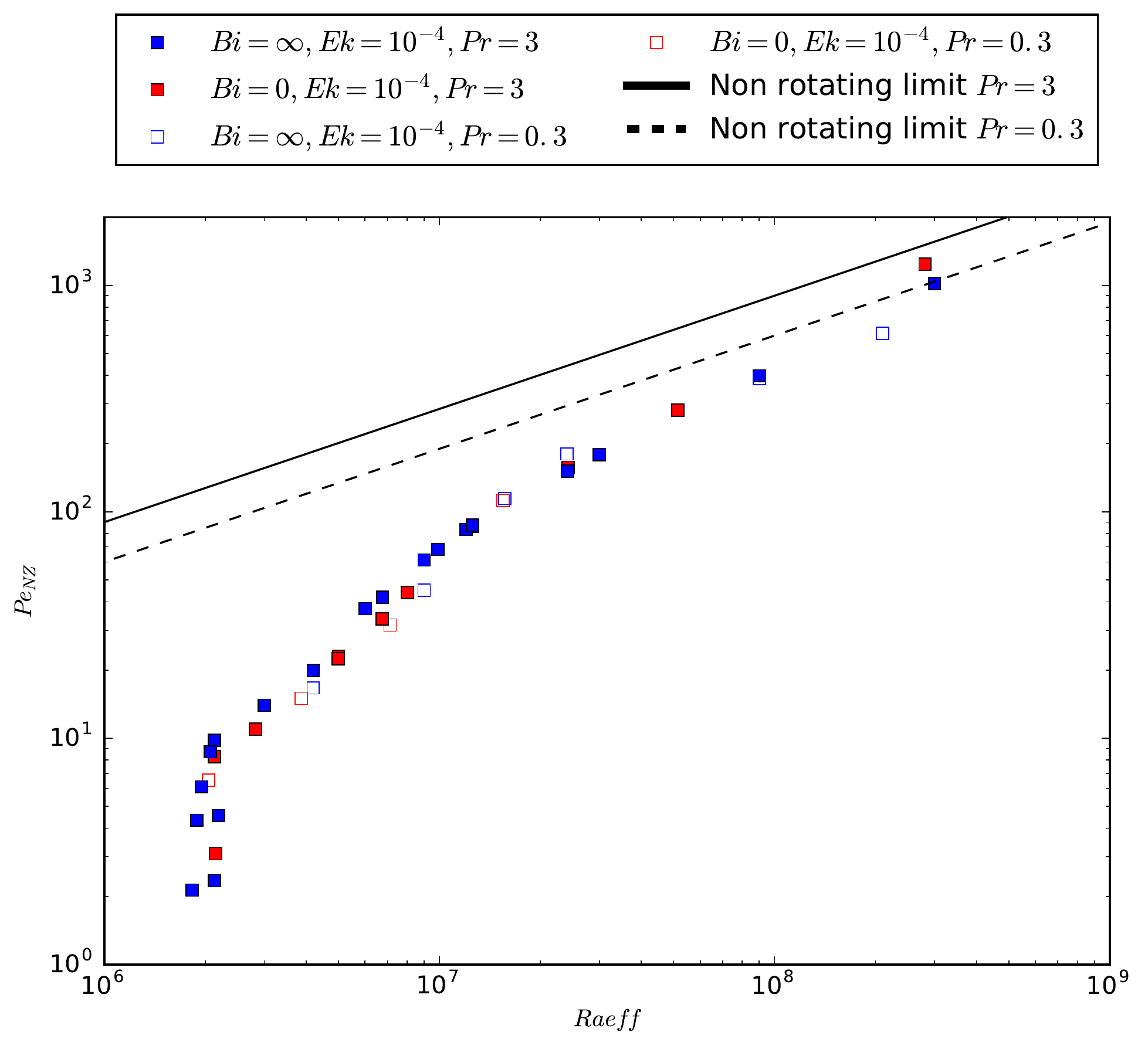}
\caption{Non-zonal component of $\Pe$ as a function of $\Ra_{eff}$ for $\Ek=10^{-4}$ cases. Similar behaviour is observed regardless of $(\Pra,\Bio)$: $\Pe$ values approaches the non-rotating limit when $\Ra_{eff}$ is increased. The overshoot observed in fig. \ref{peRa_eff} is suppressed with the subtraction of the zonal component.}
\label{peznz}
\end{figure}

We should comment on the short range of Rayleigh numbers (around $10^{8}$) for which $\Ek=10^{-4}$ cases show $\Pe$ values higher than their non-rotating counterparts, which is visible for both values of $\Pra$. This overshoot can be explained by our use of the total velocity field to compute $\Pe$. Indeed in a configuration of convection in a rotating spherical shell with upper free slip boundary condition, we expect the development of strong zonal flows \citep{yadav15}, responsible for up to $90\%$ of the total kinetic energy, in a certain range of $\Ra$. We then plot on fig. \ref{peznz} similar quantities as in fig. \ref{peRa_eff} but using only the non-zonal (i.e. $m \ne 0$ components) velocity field to compute $\Pe$. With this choice, the mentioned overshoot disappears.
Moreover, the two branches attributed to low and high $\Pra$ values for $\Ek = 10^{-4}$ which are distinct on fig. \ref{peRa_eff} collapse in fig. \ref{peznz}.
Consequently, the distinction between these two branches is probably solely due to the presence of a strong zonal component at low $\Pra$ that is absent or weaker in the $\Pra=3$ cases with $\Ra_{eff} < 10^7$. Some more comments are given about the contribution of the zonal-flows to the total kinetic energy in appendix \ref{sec:anx-zf}. 

As a conclusion from the global diagnostics we just studied ($\Nu$ and $\Pe$), for a given ($\Ek,\Pra,\Ra_{eff}$), the behaviour of thermal convection in our system is independent of the top boundary condition (FT or FF) if $\Ra_{eff} \gg \Ra_C$. Therefore, the behaviour of the system can be predicted based on results with classical boundary conditions, provided $\Ra_{eff}$ is used instead of $\Ra$.
Moreover, this $\Ra_{eff}$ can be determined \textit{a priori} in the non-rotating case from $Ra$ value. This computation can likely be extended to finite $\Ek$ values as soon as $\Ra$ is large enough to be in the weakly rotating regime. 

\subsubsection{Temperature profiles}

\label{sec:nl-Tprof}

\begin{figure}
\includegraphics[scale=0.65]{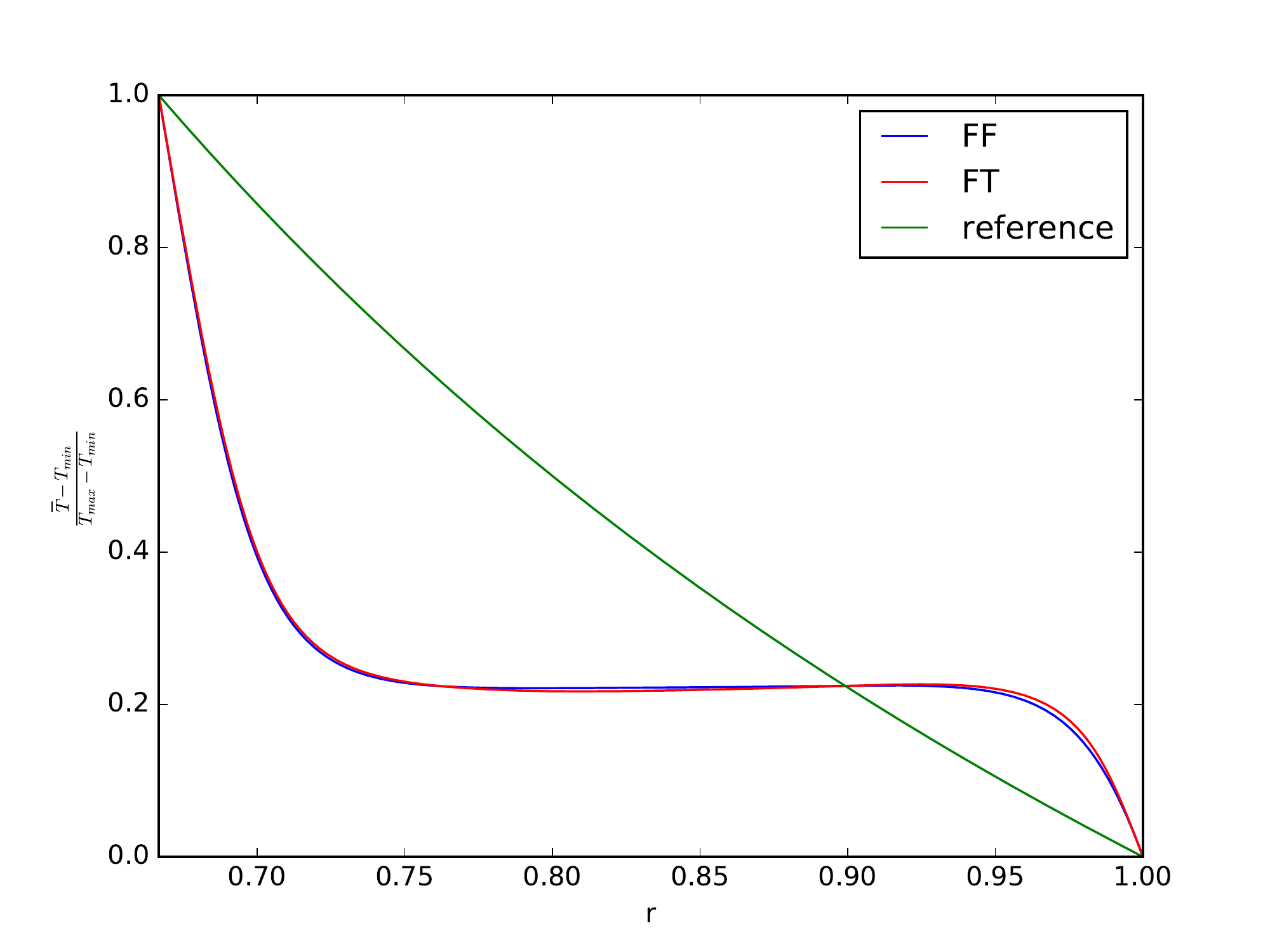}
\caption{Radial temperature profiles averaged in time and normalised by the mean temperature difference across the shell, $\frac{\overline{T}-T_{min}}{T_{max}-T_{min}}$. Reference, FF and FT profiles are compared, $\Ek=\infty,\Pra=3,\Ra_{eff} \approx 5.6\ 10^{6}$. For large enough $\Ra_{eff}$, equal $\Ra_{eff}$ configurations show similar profiles.}
\label{Tprmo}
\end{figure}

We also compare the two extreme configurations with respect to their temperature profile. We plot in fig. \ref{Tprmo} radial profiles of mean temperature of FF and FT cases characterised by the same $\Ra_{eff}$. More precisely we compute the mean radial temperature profile for several temperature fields saved at different times in the stationary state. Then, these mean radial profiles are used to produce a time averaged radial profile $\overline{T}(r)$. We finally plot $\frac{\overline{T}-T_{min}}{T_{max}-T_{min}}$ in order to compare the shape of the profiles since, by construction, $\Delta T_{eff}$ is the same in these two configurations. We can see in fig. \ref{Tprmo} that temperature profiles of FF and FT configurations are almost identical when $\Ra_{eff}$ is large enough. In the case of a weakly surcritical simulation ($\Ra \simeq \Ra_C$), temperature profiles are more similar to diffusive ones since convection is not well developed. 

\subsubsection{Flow structure}
\label{sec:flstrnl}
Convection can also be studied on the basis of the spatial structure of the flow. We consider first the structures in the meridional plane and then in equatorial ones. 

\begin{figure}
\begin{tabular}{cc}

a) \includegraphics[scale=0.2]{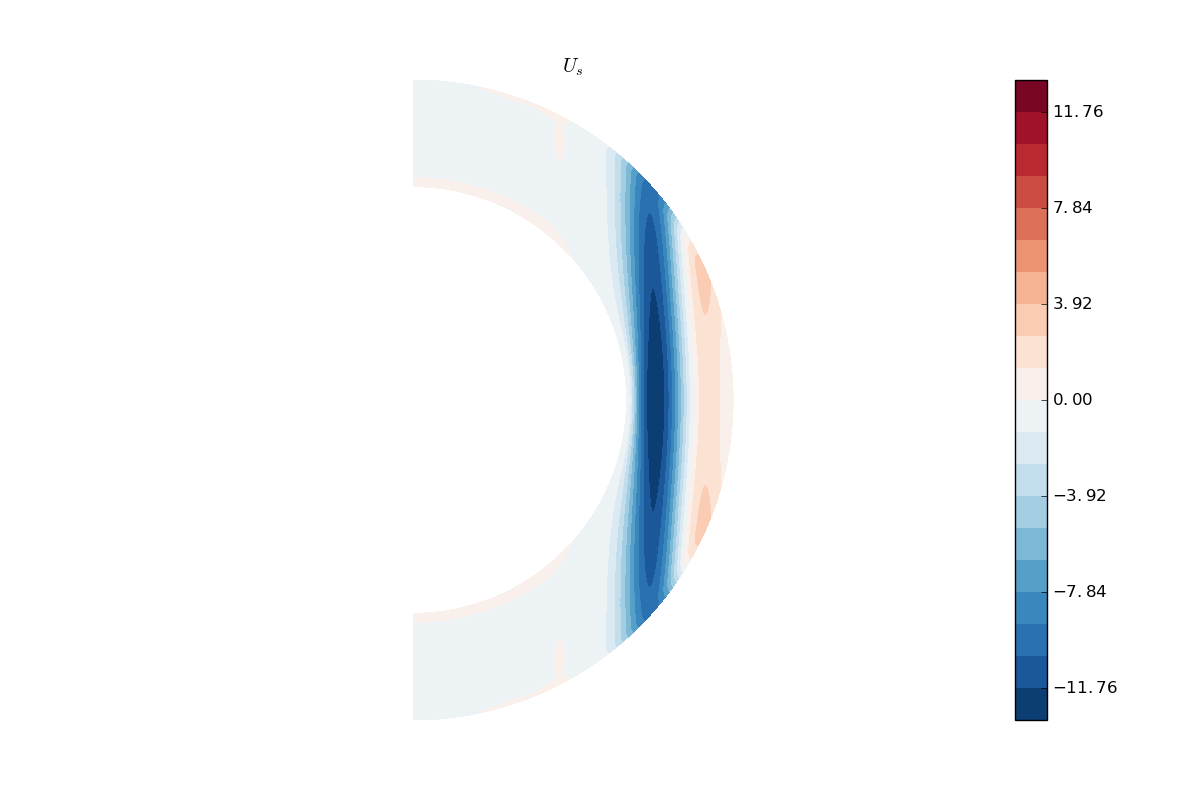} &b) \includegraphics[scale=0.2]{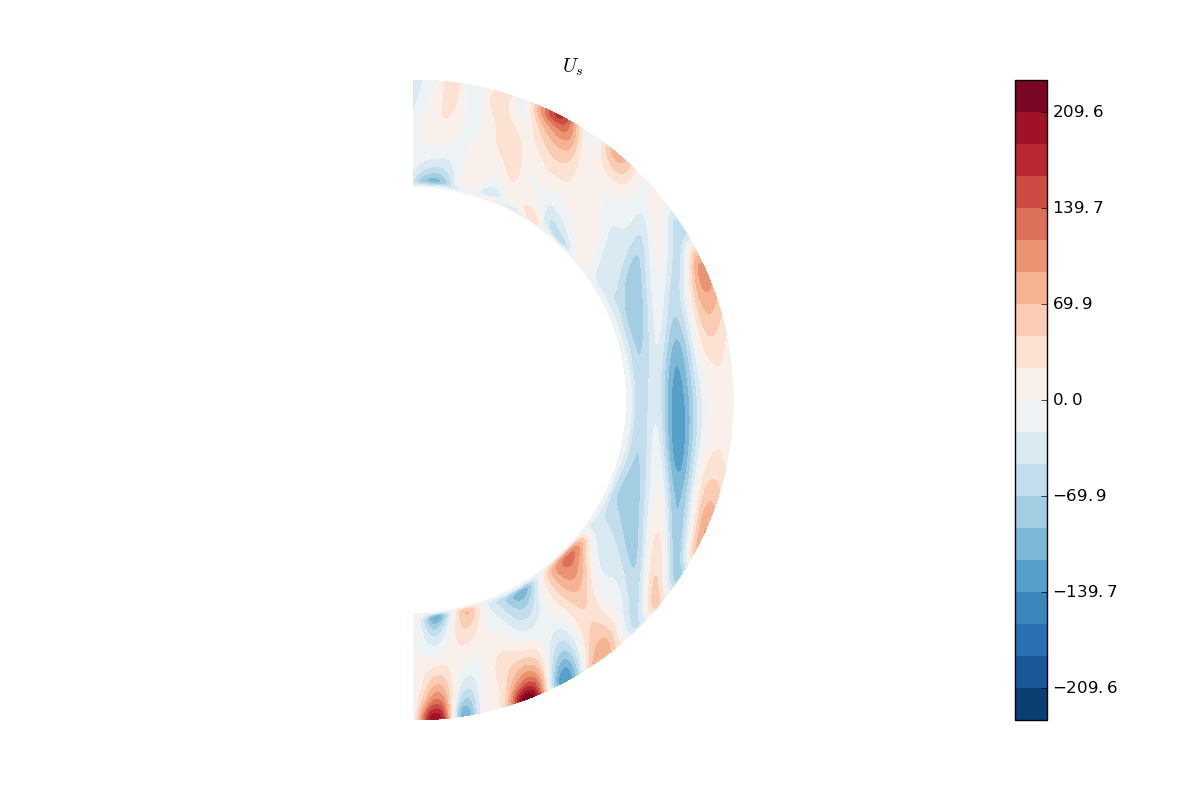}  \\ 

c) \includegraphics[scale=0.2]{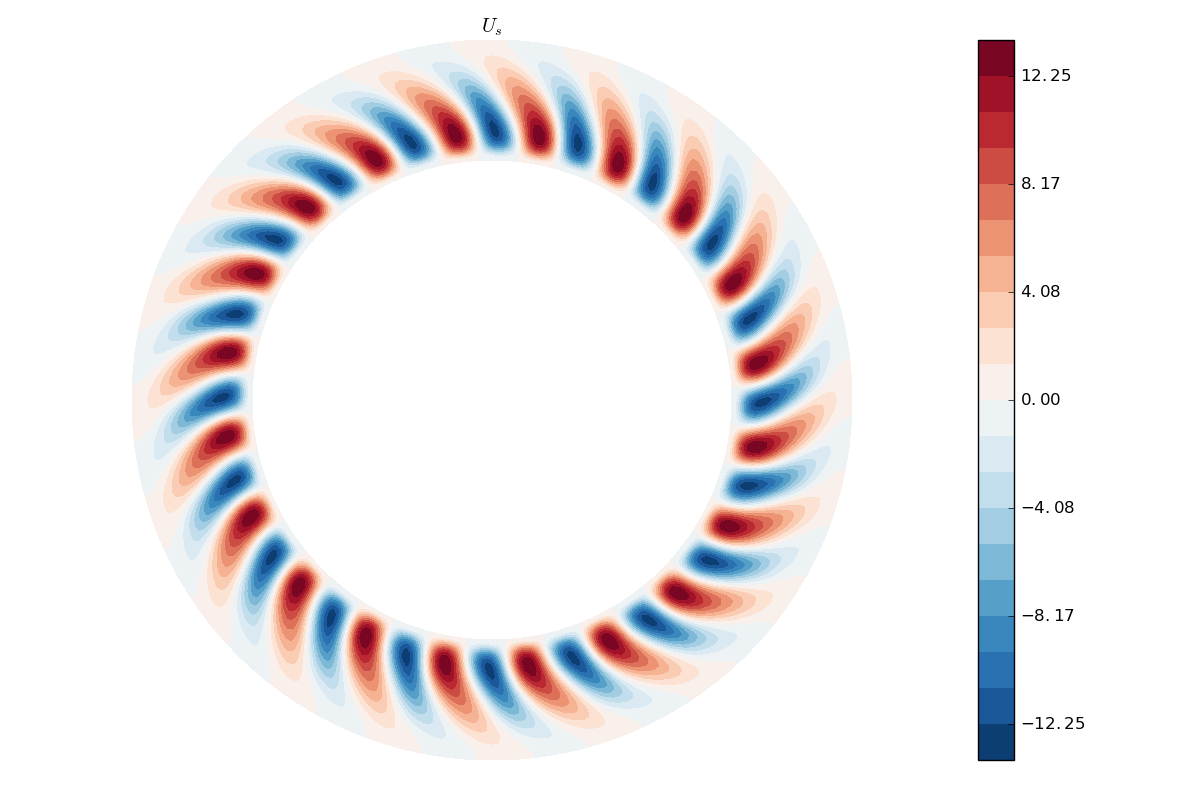} &d) \includegraphics[scale=0.2]{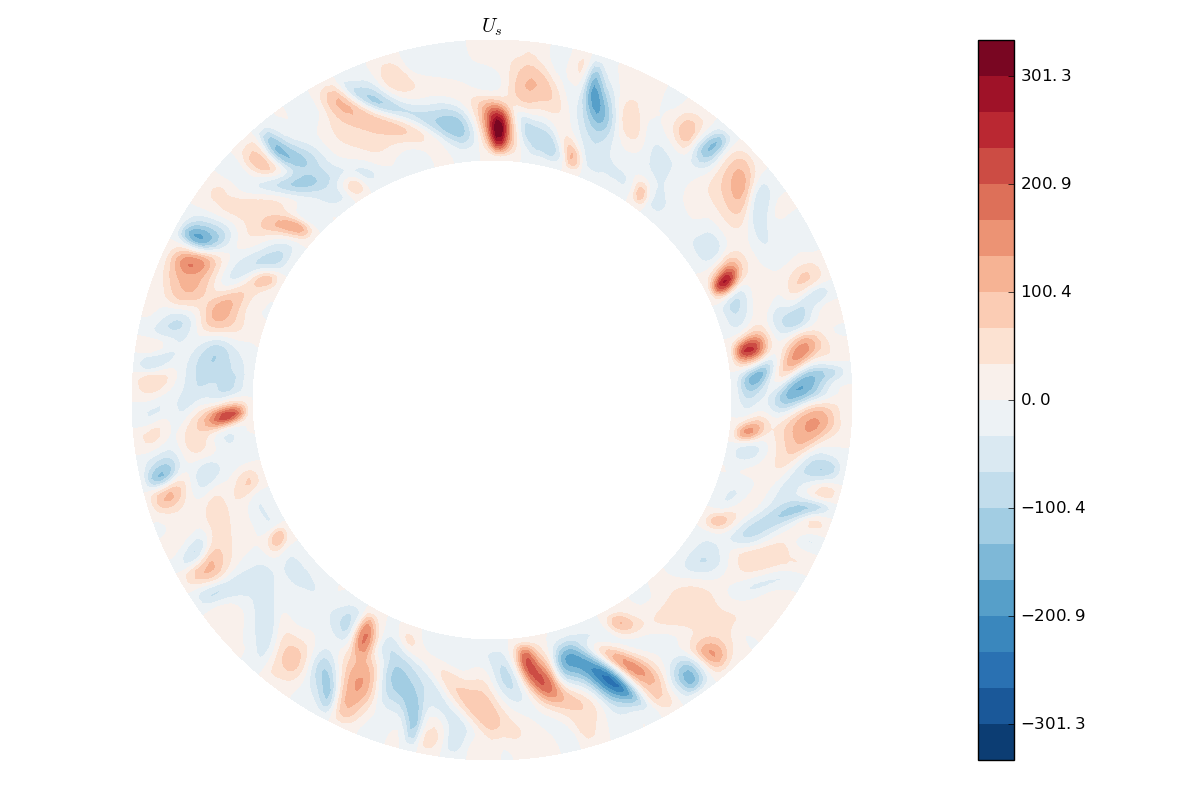}  \\ 

\end{tabular} 
\caption{Meridional (a and b) and equatorial (c and d) maps of radial velocity. a and c: $\Bio=\infty,\Ek=10^{-4},\Pra=3,\Ra=9\ 10^{6}$, b and d: $\Bio=\infty,\Ek=10^{-4},\Pra=3,\Ra=9\ 10^{7}$. a: the flow is only active outside the tangent cylinder, b: the flow is active in the whole shell. c and d: flow loses its lateral organization when $\Ra$ is increased of a factor 10.}
\label{mervel}
\end{figure}

The meridional slices enable us to investigate the disposition of the flow with respect to the tangent cylinder in the rotating cases (without rotation this concept is not relevant). As shown in fig. \ref{mervel}(\textit{a}) moderate supercriticality implies that flow is not active in the tangent cylinder. When $\Ra$ is increased enough (\textit{b}), this spatial organization is lost and the flow occupies the whole shell. Nevertheless in both cases, the influence of rotation can be noticed thanks to the alignment of the flow along columns of axis $\mathbf{\Omega}$.

As a remark, we noticed in some cases (e.g. $\Bio=\infty,\Ek=10^{-4},\Pra=3,\Ra=3\times 10^{7}$ i.e. in rotating cases at moderate $\Ra/ \Ra_C$) the development of an anti-symmetric component of the velocity field with respect to the equatorial plane. This component was never stronger than the symmetric one. The anti-symmetric component of the flow is mostly seen within the tangent cylinder where the Taylor-Proudman constraint does not link both hemispheres.

Then, we choose to compare the respective properties of structures at equal $\Ra / \Ra_{C}$ in order to compare slightly supercritical configurations. For higher $\Ra$, equal $\Ra_{eff}$ plots are compared since it has been shown that this number was an accurate criterion to fairly compare FF and FT configurations.

\begin{figure}
\centering
\begin{tabular}{cc}
\hline 

a) \includegraphics[scale=0.2]{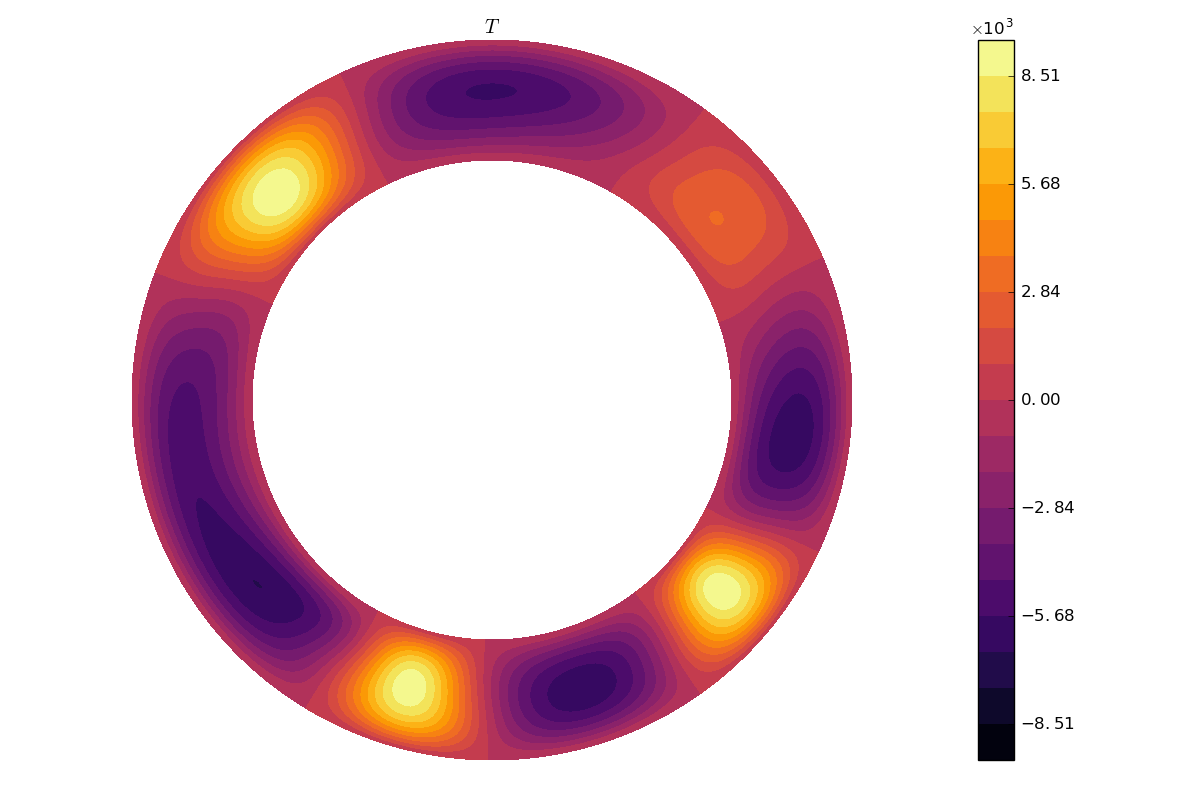} & b)\includegraphics[scale=0.2]{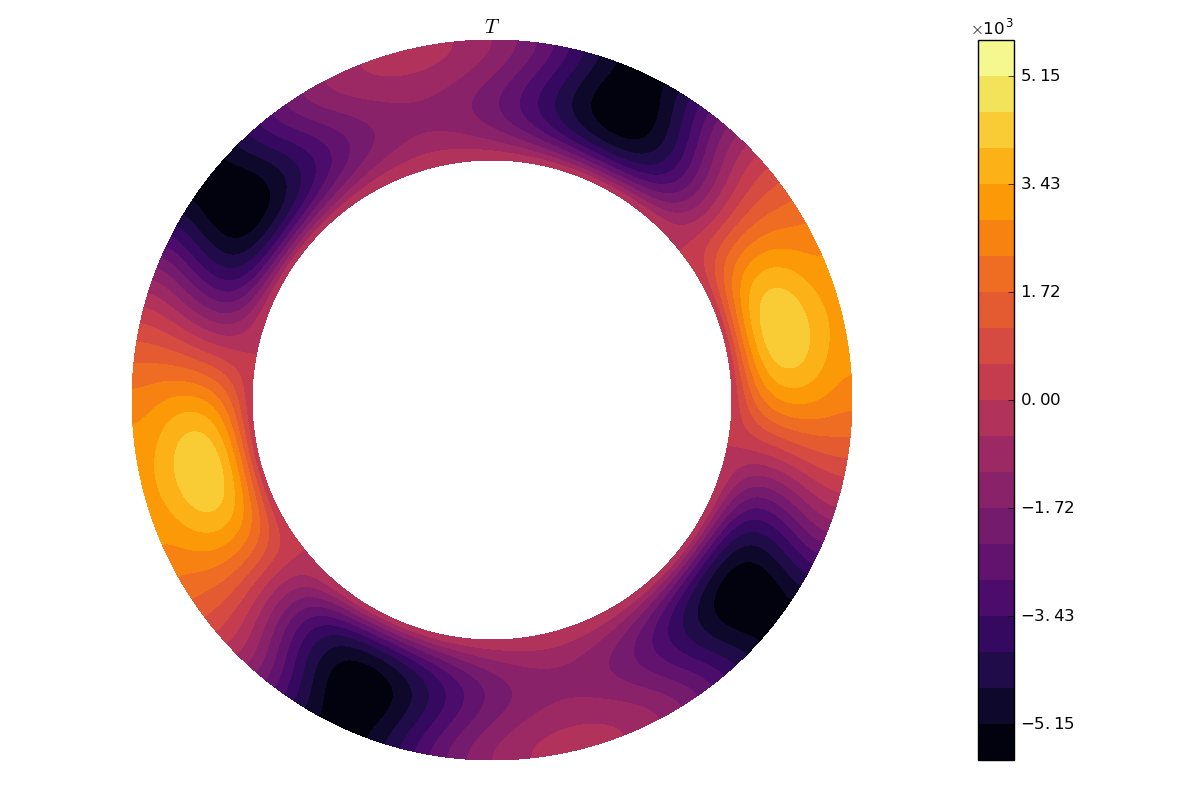} \\

c) \includegraphics[scale=0.2]{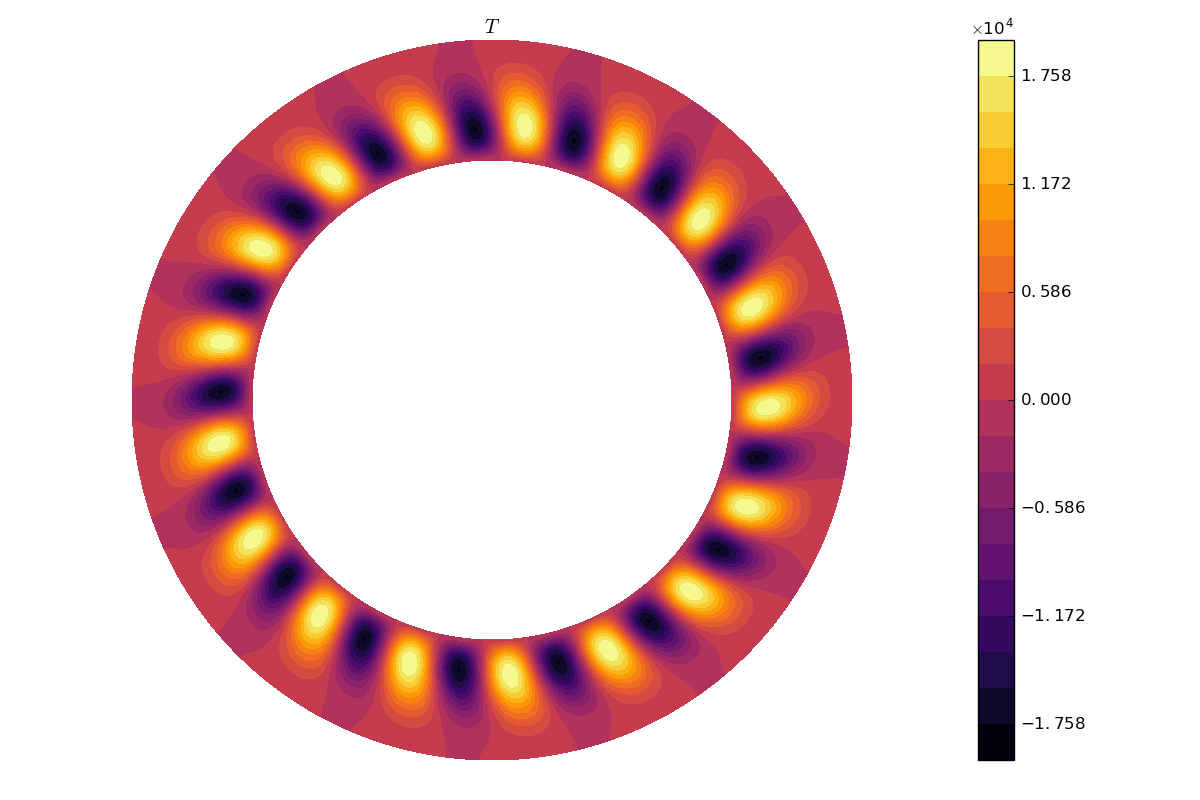} & d) \includegraphics[scale=0.2]{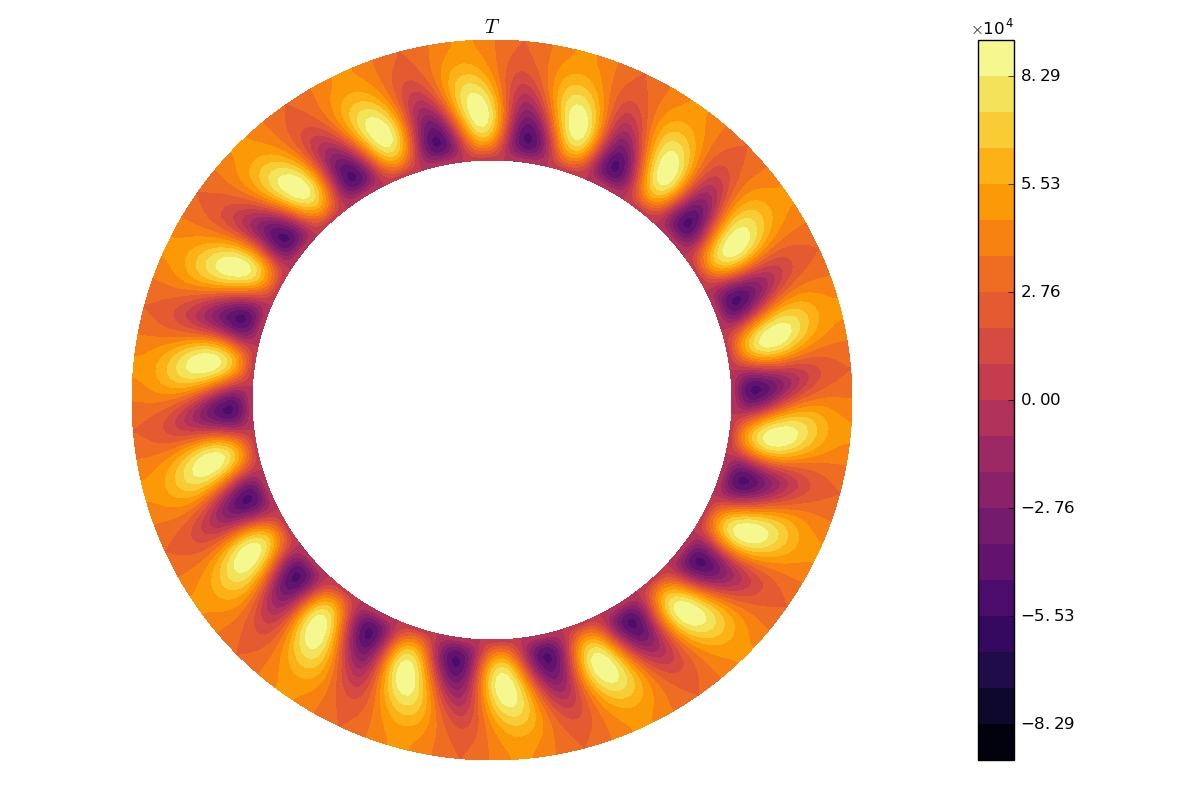} \\ 

e) \includegraphics[scale=0.2]{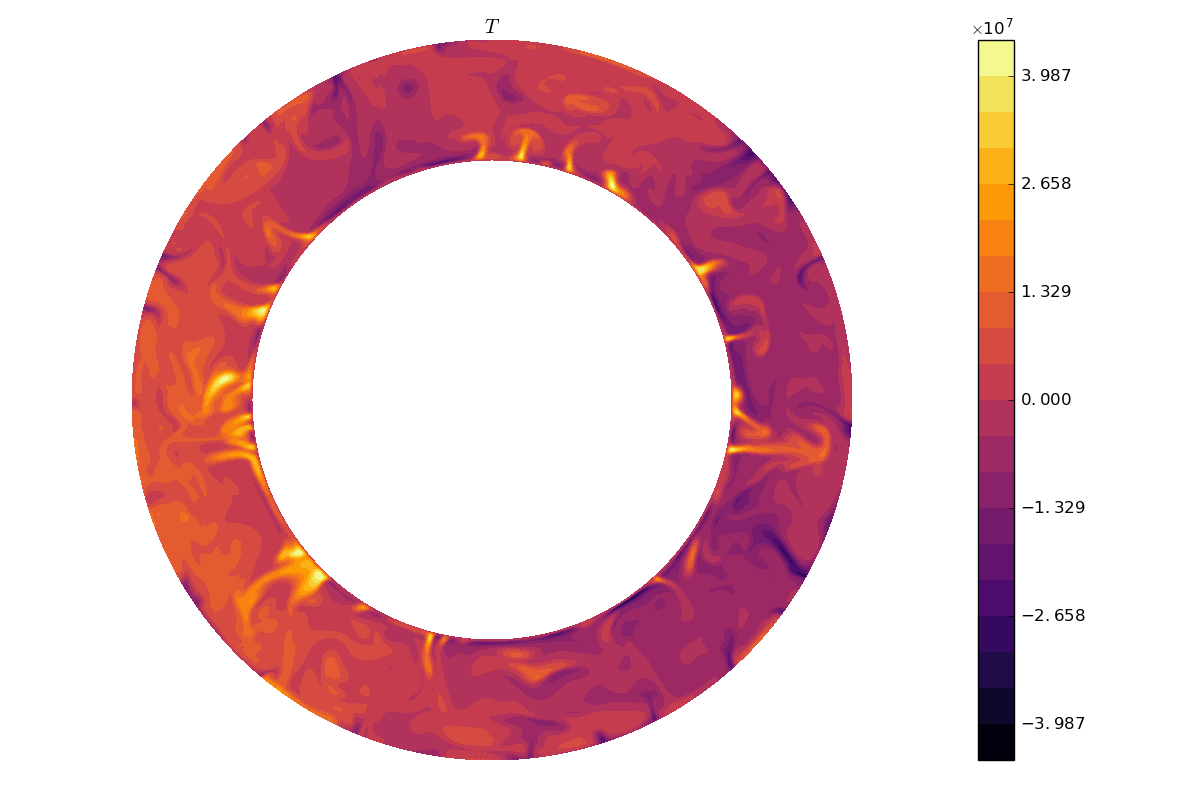} & f) \includegraphics[scale=0.2]{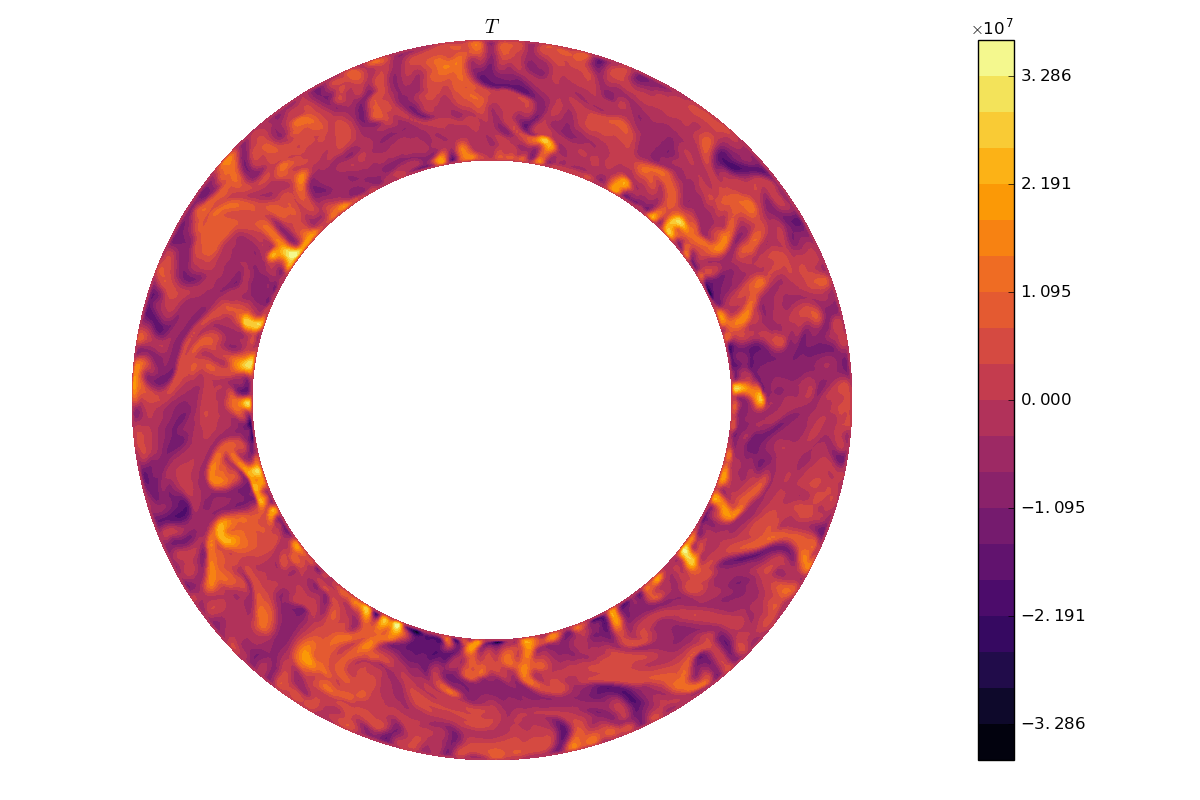} \\ 

g) \includegraphics[scale=0.2]{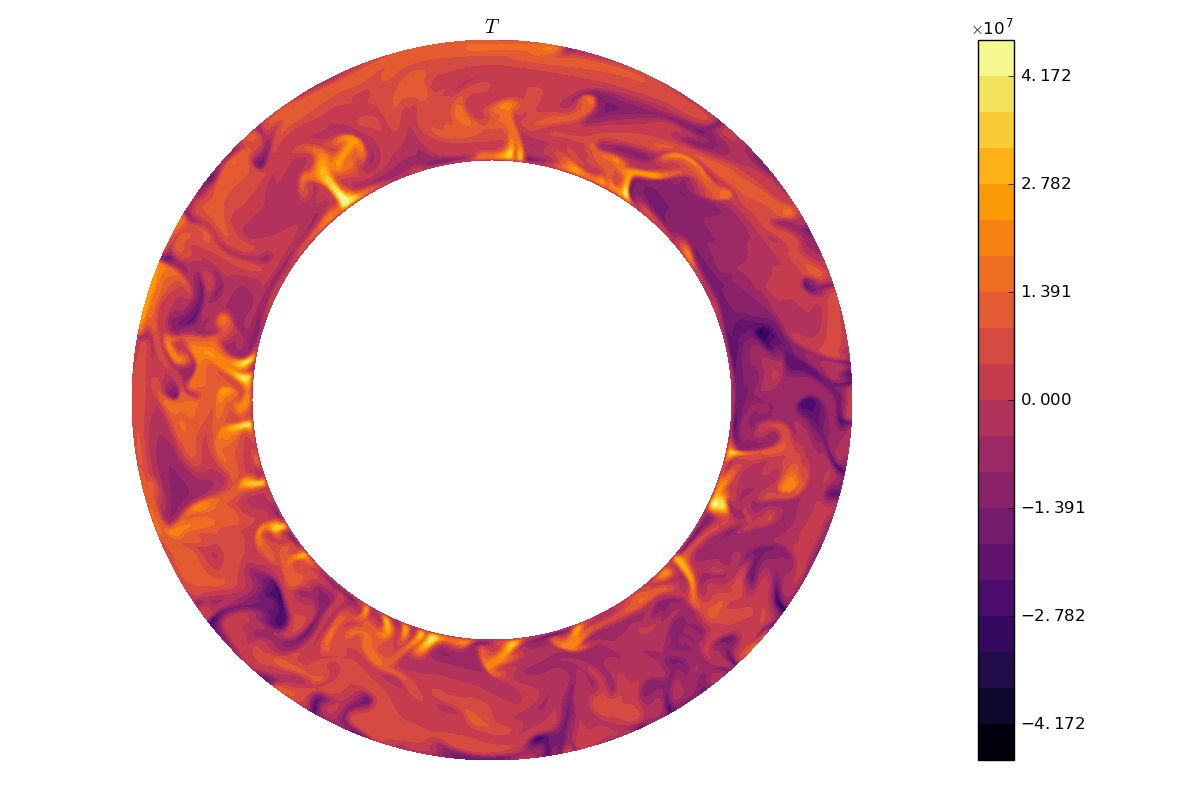} & h) \includegraphics[scale=0.2]{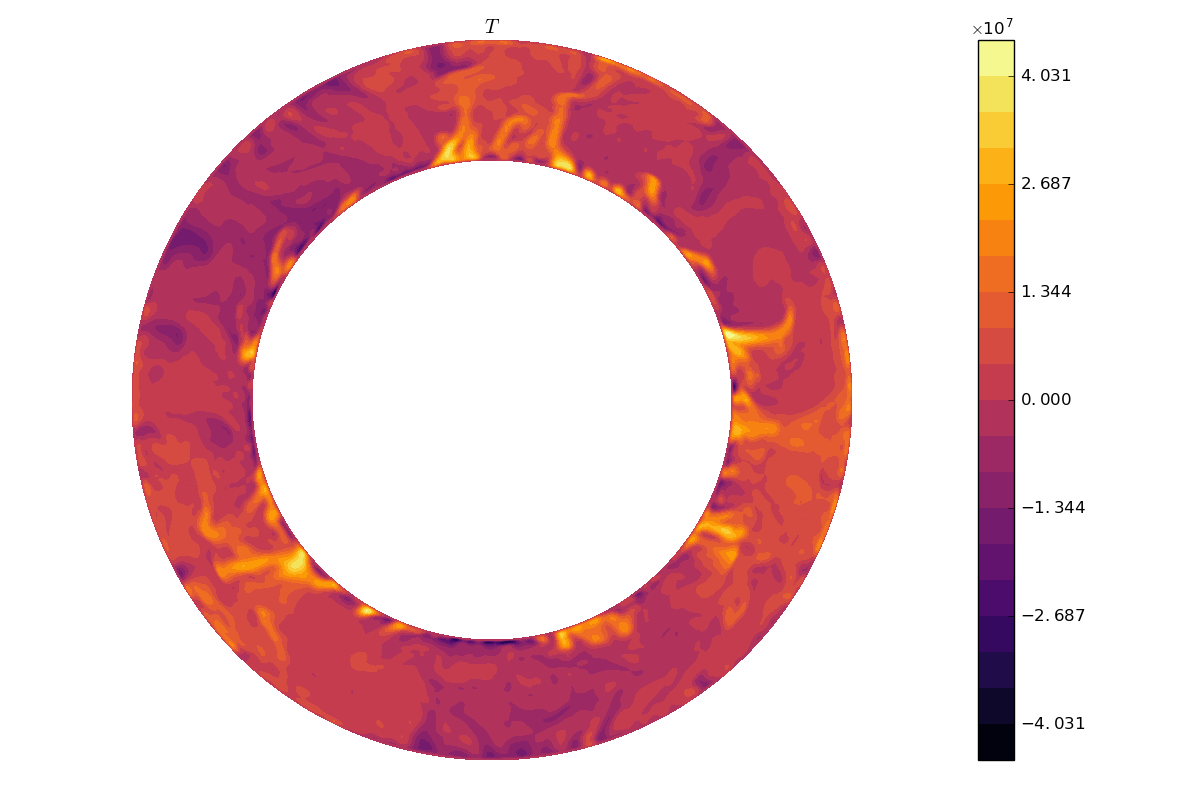} \\ 
 
\end{tabular} 
\caption{\footnotesize{Equatorial maps of temperature anomalies ($l=0$ component is subtracted, $\Pra=3$ for all cases); left $\Bio=0$ (FF), right $\Bio=\infty$ (FT); c,d,g,h: rotating cases ($\Ek=10^{-4}$), a,b,e,f: non rotating ones; a,b,c and d: $\Ra/ \Ra_{C} \simeq 1.1$, e,f,g and h: $\Ra_{eff}=3.10^{8}$. Except difference of thermal upper boundary layer, structures of FF and FT thermal anomalies are similar for each line.} }
\label{eqtemp}
\end{figure}

We can see on fig. \ref{eqtemp} that it is usually possible to discriminate FF from FT cases without consideration of ($\Ek,\Pra$) by checking if temperature anomalies extend to the upper boundary or not. This is illustrated clearly in the non-rotating configuration (see first line for slightly surcritical case and third one for higher $\Ra / \Ra_C$) and in the rotating case (see second line for slightly surcritical simulations and last one for moderate $\Ra / \Ra_C$). 

Except for the vertical extent of the thermal anomalies near the upper boundary, it is nearly impossible to visually distinguish between FF and FT equatorial temperature snapshots on the basis of the size of the anomalies or their spatial organization. We extend this comparison to 3D structures in the appendix \ref{sec:anx-3D}.

In the slightly supercritical rotating cases (second line) we observe the lateral periodicity ($m=17$) predicted by the linear stability analysis (see fig. \ref{dphas} and \ref{delm}), which is identical for both FF and FT boundary condition. It is worth noting to remark the high temperature equatorial anomaly which is characteristics of the fixed flux configuration.

\subsection{FF to FT transition}
\label{sec:nlbiot}
We now focus on the transition from FF to FT configurations produced by an increase of the Biot Number from 0 to infinite limit values. We restrict our set of parameters to the $\Pra=3$ simulations. For various $\Ra$ and $\Ek$ we determined $\Nu$ when the Biot number evolves from one limit to the other.

\subsubsection{Transition with respect to $\Nu$}
\label{sec:nlbinu}
\begin{figure}
\includegraphics[scale=0.65]{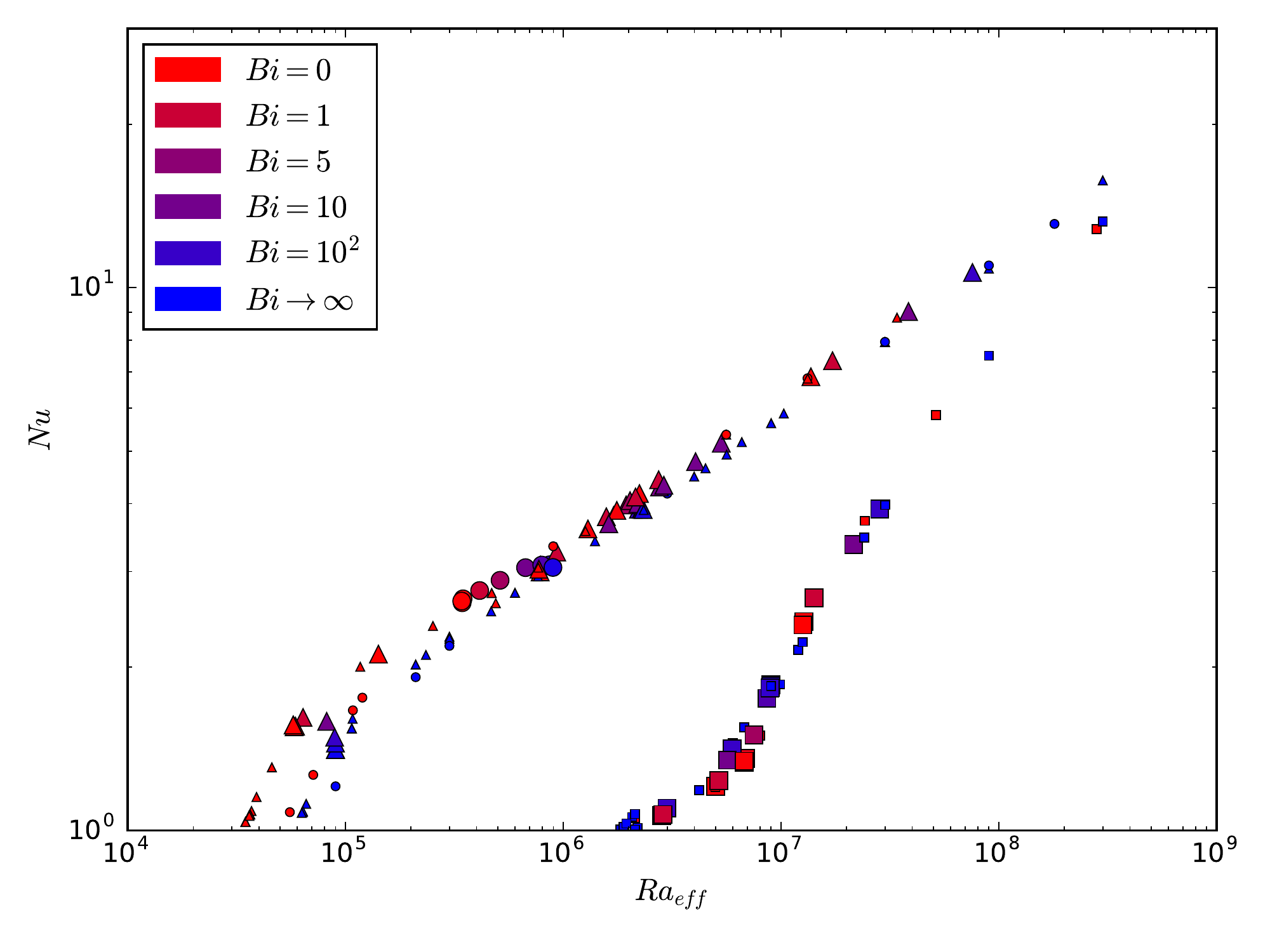}
\caption{Nusselt number $\Nu$ as a function of $\Ra_{eff}$ for the same representation code as in fig. \ref{nu}. Intermediate $\Bio$ cases have been added to the fig. \ref{nu} data. They are plotted with big symbols, coloured according to their $\Bio$ value. As in the fig. \ref{nu}, points are distributed along two main branches determined by the value of $\Ek$. Except for ($\Ek= \infty$, $\Ra \approx \Ra_{C}$) points, intermediate $\Bio$ points are located on the same branches as extremal ones and then follow the same scalings as described in fig.~\ref{nu}.}
\label{Raeff-bi}
\end{figure}

Intermediate $\Bio$ values cases can be compared with FF and FT ones on the basis of their $\Nu$, as plotted on fig. \ref{Raeff-bi} as a function of $Ra_{eff}$. They follow the same behaviour as for FF and FT cases, which shows that $Ra_{eff}$ constitutes the relevant variable characterizing this different types of thermal convection subjected to different boundary conditions. We did similar observations on the basis of $\Pe$ rather than $\Nu$.

As in the section \ref{compglob} we compute $\Ra_{eff}$ as a function of $\Bio$ for $(\Ek=\infty,\Pra =3)$. We found that for fixed $\Ra$, $\Ra_{eff}$ increases continuously and monotonically from FF to FT values following
\begin{equation}
    \Ra_{eff}(\Bio)=\Ra_{eff,FF}+\Delta \Ra_{eff} \frac{\Bio}{\Bio+\Bio_{t}},
    \label{modraeff}
\end{equation}
with $\Delta \Ra_{eff}$ the difference between FF and FT values of $\Ra_{eff}$ (which can be determined by eq. \ref{ffraeff}) and $\Bio_{t}$ a threshold of approximate value 10 -- we noticed that $\Bio_{t}$ increased slightly with $\Ra$.
Our data show a good agreement with this model on three orders of magnitude of $\Ra$ (discrepancies are always smaller than $10\%$).

\subsubsection{Transition for a given effective Rayleigh number}

\begin{figure}
   \centering
    \includegraphics[scale=0.65]{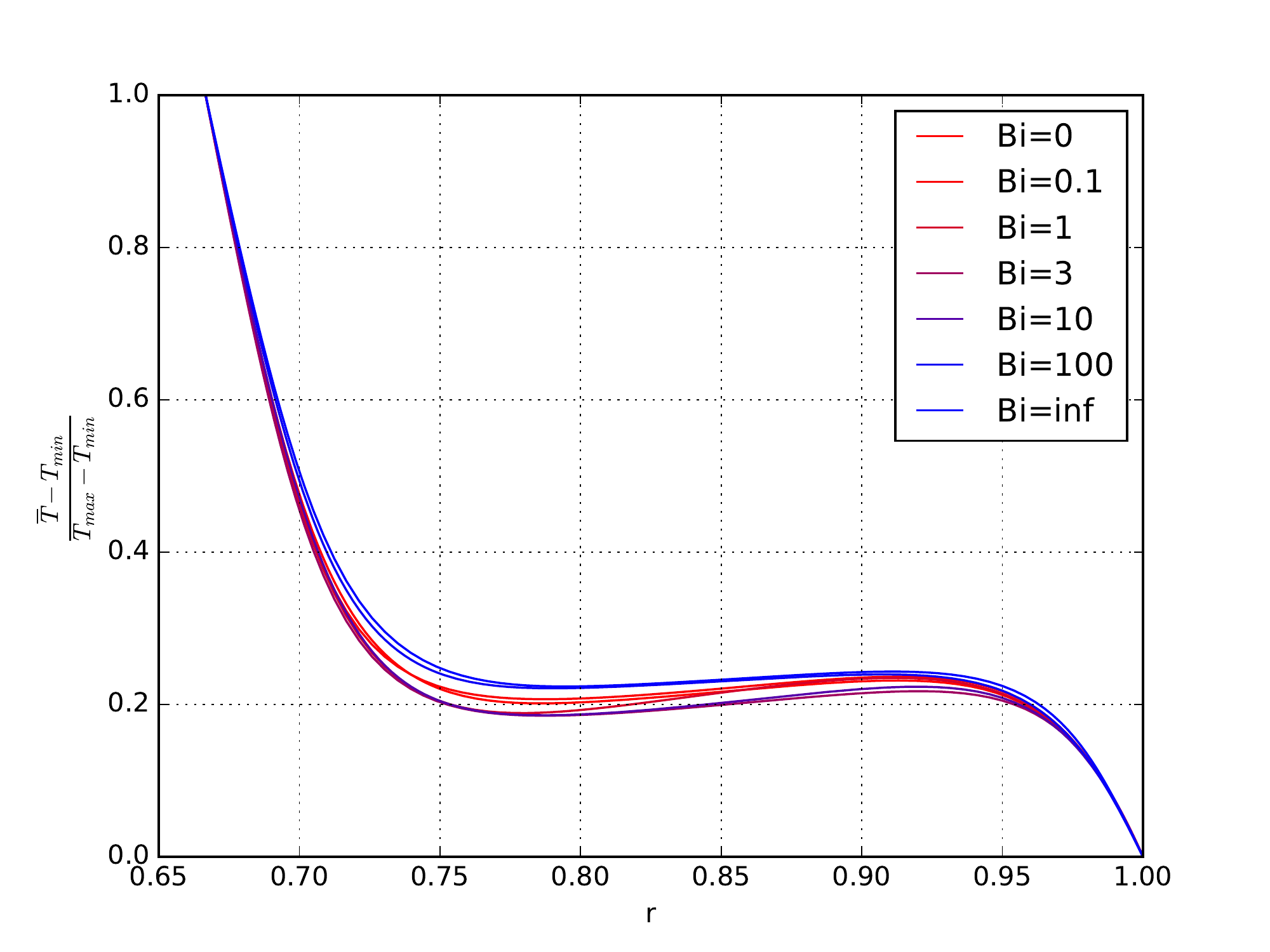}    \caption{Normalised and time-averaged radial temperature profiles for quasi constant $\Ra_{eff} \simeq 2.2 \times 10^{6}$ but variable $(\Ra,\Bio)$, ($\Ek=\infty,\Pra=0.3$). Similar profiles are obtained, nevertheless, three different behaviour can be distinguished: for large (FT), small (FF) and intermediate $\Bio$ values. Exact values of $(\Ra_{eff},\Ra)$ for each $\Bio$ are: $\Bio=0,\Ra_{eff}=2.18\times 10^{6}, \Ra=9\times 10^{6}$; $\Bio=0.1,Ra_{eff}=2.24\times 10^{6}, \Ra=9\times 10^{6}$; $\Bio=1,\Ra_{eff}=2.14\times 10^{6}, \Ra=6.6\times 10^{6}$; $\Bio=3,\Ra_{eff}=2\times 10^{6}, \Ra=4.5\times 10^{6}$; $\Bio=10,\Ra_{eff}=2.2\times 10^{6}, \Ra=3.3\times 10^{6}$; $\Bio=100,\Ra_{eff}=2.2\times 10^{6}, \Ra=2.34\times 10^{6}$; $\Bio=\infty,Ra_{eff}=2.18\times 10^{6}, \Ra=2.34\times 10^{6}$.}
  \label{radprofbi}
\end{figure}

Even if equal $\Ra_{eff}$ implies similar $\Nu$ regardless of $\Bio$, is it possible to detect differences in the flow structure at constant $\Ra_{eff}$ with respect to the $\Bio$ value?
As an illustration we plot the radial profiles of temperature (fig. \ref{radprofbi}) of seven cases of approximately equal $\Ra_{eff}$ and variable $\Bio$. Note that we could not obtain a collection of cases with exactly constant $\Ra_{eff}$ but variable $(\Ra,\Bio)$, since the value of $\Bio_{t}$ used in eq. \ref{modraeff} evolves slightly with $\Ra$, nevertheless differences are below $10 \%$.
These profiles have been time averaged over fields extracted from the statistically stationary regime. Here we focused on non-rotating systems with $\Ra \gg \Ra_C$.
As one can see, the general structure of the profiles is the same: the temperature difference is concentrated in narrow boundary layers located at both shell limits, while the interior of the shell is homogenised by convection and exhibits a slightly stabilizing temperature profile.
While the profiles are nearly identical near the boundaries, the middle temperature is affected by $\Bio$. This dependency is not monotonous: intermediate (i.e. between 1 and 30) $\Bio$ profiles have the lowest mean temperature, FF ones show medium temperature and FT cases the highest mean temperature.

\begin{figure}
    \centering
    \includegraphics[scale=0.65]{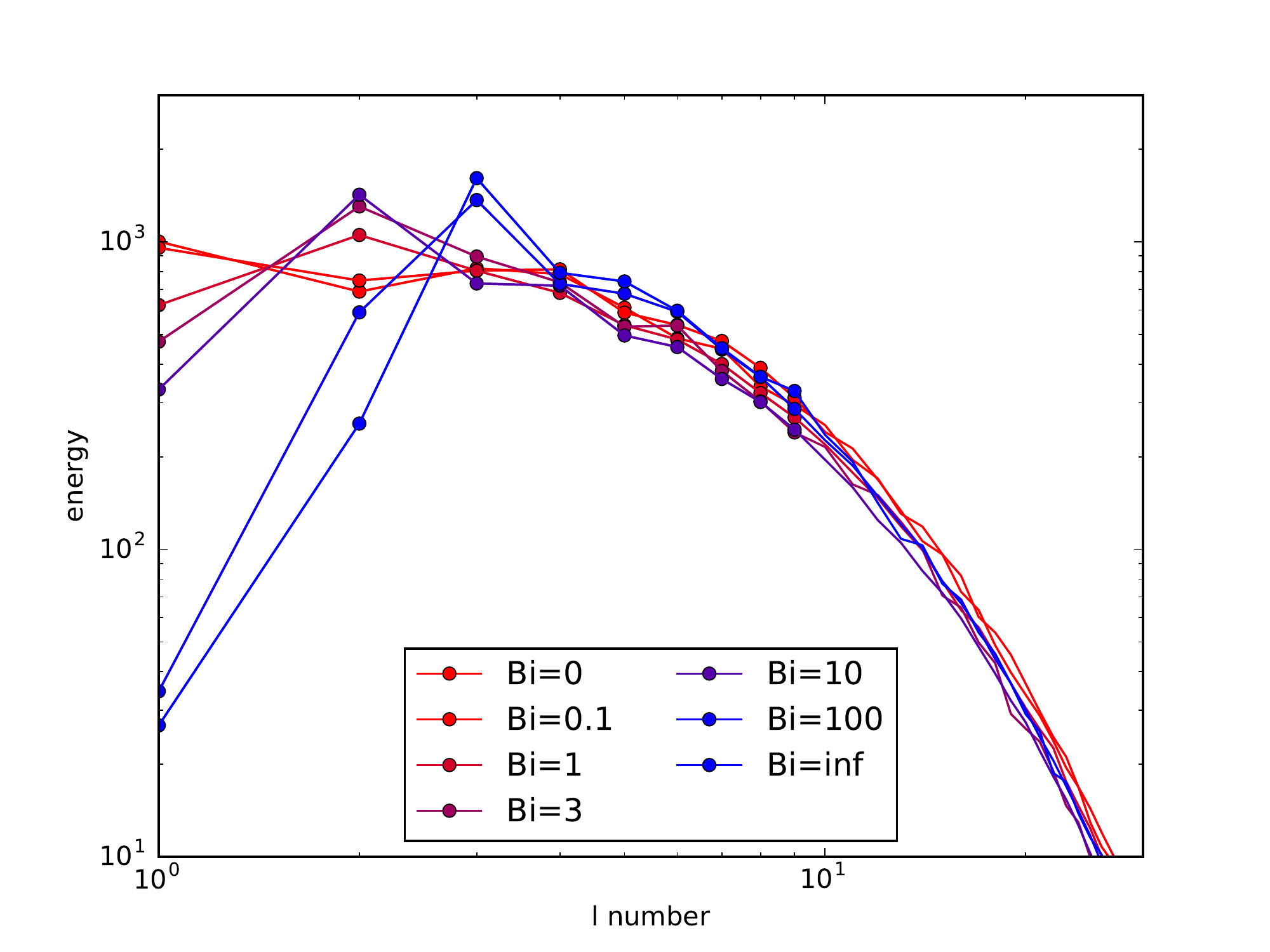}
    \caption{Time average energy spectrum of the velocity (sum of toroidal and poloidal components over the whole shell) for roughly constant $\Ra_{eff}\simeq 2.2 \times 10^{6}$, $\Ek=\infty$, $\Pra=3$ but variable $\Bio$. For $\Bio\le 0.1$, 
    $l=1$ is the dominant mode, 
    intermediate values ($1\le\Bio < 10$) favour $l=2$, and $l=3$ 
   dominates at high $\Bio$ 
   like in FT configuration. The total 
   kinetic energy in all these 
    configurations varies by 
    less than 10$\%$.
    }
    \label{specU}
\end{figure}

In order to give a quantitative description of the spatial structure of the flow, we plotted the velocity spectrum (fig. \ref{specU}) for each of theses cases. With the information provided by the spectra it is possible to identify the three regimes suggested by the radial profiles of the fig. \ref{radprofbi}. Indeed, the low, intermediate and high $\Bio$ value cases are characterised respectively by a dominance of $l=1,2$ and 3. Finally small-scale flows are favoured at higher $\Bio$ (that is FT-like top boundary condition) while the different spectra become indistinguishable as soon as $l \ge 7$.

To conclude this section, we have shown that even if $\Ra_{eff}$ is the relevant parameter to compare convection efficiency of FF and FT cases, differences in the flow structure are observed at constant $\Ra_{eff}$ but variable $\Bio$ (without rotation). For rotating configurations (particularly at $\Ek=10^{-4}$), it has been impossible to bring out important differences of the flow structure between FF and FT cases.

\subsubsection{Convection regimes for variable Biot number}

\begin{figure}
    \centering
    \includegraphics[scale=0.65]{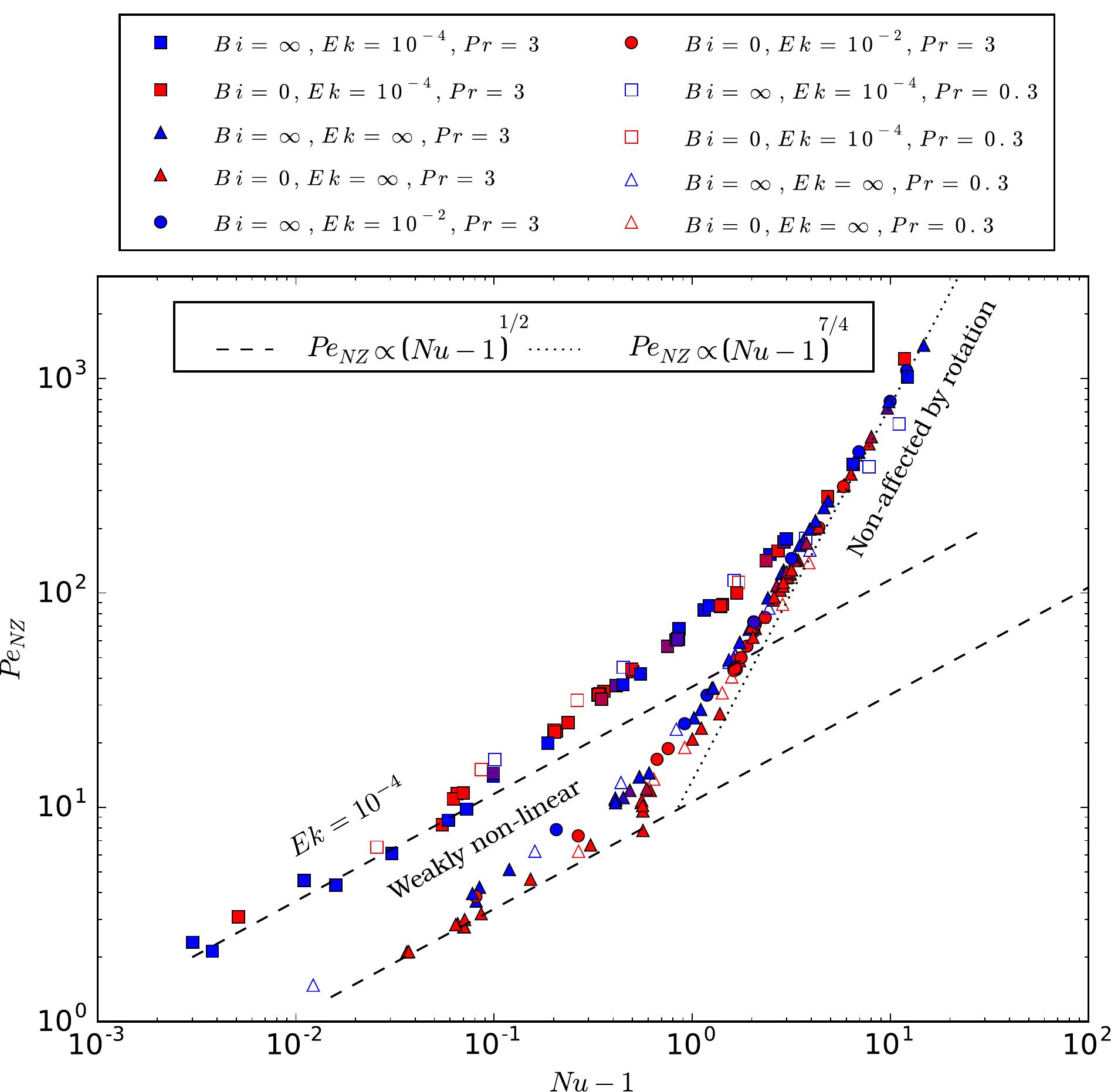}
    \caption{Non-zonal component of $\Pe$ as a function of $\Nu-1$ for various $\Ek,\Pra,\Bio$ coded by shape, filling and colour of the symbols respectively (as in fig. \ref{Raeff-bi} and \ref{nu}). We distinguish two main regimes: the weakly non-linear regimes at low $\Pe$ and $\Nu\simeq 1$ following the two broken-line models and the weakly or non-rotating regimes represented by the dotted line.}
    \label{penu}
\end{figure}

Rotating and non-rotating thermal convection are known to show various regimes depending on the intensity of the thermal forcing \citep{gastine16, long2020}. Usually these successive regimes are spotlit by scaling laws deduced from $\Nu=f(\Ra)$ or $\Pe=f(\Ra)$ graphs.
Here we choose to combine these two in a $\Pe_{NZ}=f(\Nu-1)$ graph (more usual graphs have been presented in the beginning of this section \ref{sec:extrnl}, see fig. \ref{peRa_eff} and \ref{Ra_eff}). The non-zonal component of $\Pe$ is solely used since its exclude zonal winds that do not participate to the radial thermal transfer. We observe on fig. \ref{penu} different branches on which points are distributed mainly following their Ekman number ($\Ek=10^{-2}$ points are poorly distinguishable from non-rotating ones). For low $\Pe$ and $\Nu$, two branches of specific $\Ek$ are visible. They are well modelled by a $\Pe\propto (\Nu-1)^{1/2}$ relation. For higher $\Nu$, these two branches join a new one with a steeper slope ($\Pe \propto (\Nu-1)^{7/4}$).
The non-rotating branch joins this steeper branch for $\Nu\simeq 1$ while the rotating branch joins it for $\Nu \simeq 20$. We observe no strong influence of $\Bio$ or $\Ek$ on these branches except in the non-rotating one for small $\Nu-1$: FT points seem to show systematically slightly higher $\Pe$ compare to FF ones.
Remarkably, there is almost no dependence on $\Pra$ in this representation -- although only $\Pra=0.3$ and 3 are considered.

We further interpret these different branches relatively to convection regimes. 
Small $\Pe$ and $\Nu-1$ are characteristic of a slightly surciritical convection that is a weakly non-linear (WNL) regime in which the convective flow is mainly monitored by $\frac{\Ra}{\Ra_C}$. From usual scaling laws (see \citet{busse1986} and \citet{gillet2006}) relative to the WNL regime, we expect that on the one hand \begin{equation}
\label{eq:Nu-wnl}
\Nu-1\propto \left(\frac{\Ra}{\Ra_{C}}-1\right),
\end{equation}
and on the other hand, when the mean thermal profile remains similar to the diffusive one (which is appropriate for $\Ra \simeq \Ra_C$), we expect following the equilibrium between viscous dissipation rate and mean buoyancy power (see eq. $3.3$ in \citet{gastine16} and eq. $2.16$ in \citet{Gastine2015}) 
\begin{equation}
\frac{\Ra(\Nu-1)}{\Pra^{2}} \propto \frac{\nu U_{C}^{2}}{\mathcal{L}_{C}^{2}},
\end{equation}
where $U_{C}$ and $\mathcal{L}$ are the respective typical velocity and length of the convective flow.
Keeping only terms of order 1 in $\Nu-1\rightarrow 0$ leads to
\begin{equation}
\Pe\propto ((\Nu-1)\Ra_{C})^{1/2} \mathcal{L}_C.
\label{relseuil}
\end{equation}

This dependence ($Pe_{NZ}\propto (\Nu-1)^{1/2}$) is illustrated by the broken lines of the fig. \ref{penu}. Since $\Ek$ has a strong influence on the $\Ra_C$ as well as on $\mathcal{L}_{C}$ (see section \ref{sec:lin-stab}) --  decreasing $\Ek$ makes the former increase and the latter decrease -- the separation of the points between a rotating and a non-rotating branch can be understood. As shown in the linear stability analysis section \ref{sec:lin-stab}, rotation impacts poorly the onset of convection for $\Ek>10^{-3}$, which explains why results for $\Ek=10^{-2}$ fall on the non-rotating branch. Since both $\Ra_C$ and $\mathcal{L}_C$ can be expressed as $\Ek$ functions using appropriate scaling laws, a $\Pe \propto \Ek^{-1/2}$ dependency can be predicted. Nevertheless we are not able to test it since the scaling laws are established for smaller $\Ek$ than those we used here. Finally, $\Bio$ has generally a weak or no influence on $\Ra_C$ except in the non-rotating configuration which explains the small difference between FF and FT points in the non-rotating branch. 

The highest values of $\Pe,\Nu$ are characteristic of the non- or weakly rotating regime. 
We presented above (\S~\ref{sec:nlgdiag}) the usual scaling laws used in the non-rotating limit for $\Pe$ and $\Nu$ (see fig. \ref{peRa_eff} and \ref{nu}). The combination of both these laws leads to
\begin{equation}
    \Pe\propto \Nu^{7/4},
    \label{relnr}
\end{equation}
which is the slope observed in fig. \ref{penu}, for high $\Nu-1\simeq \Nu$.
Note that our scaling law is very similar to the one found for this regime in another study of thermal convection with a different system and mechanical boundary conditions \citep{king2013}.
In the non-rotating branch, $\Ek$ is no longer a discriminating factor as expected.
Remarkably, lower $\Pra$ points are not easily distinguished from the $\Pra=3$ group of points. This is worth noting since both $\Pe$ and $\Nu$ exhibit dependencies on $\Pra$ (see Figs. \ref{Ra_eff} and \ref{peRa_eff}).

Between these WNL and non-rotating regimes, a transitional regime is expected for rotating simulations. We observe it roughly between $\Nu\simeq 1.1$ and $\Nu\simeq 20$, it is characterised by an intermediate slope between the two end-members. Following a former study \citep{gastine16} we expect a somehow higher ($\Nu \simeq 1.5$) value of $\Nu$ as a lower limit of this regime. Note that since we never use $\Ek$ smaller than $10^{-4}$, we do not expect to significantly observe a rapidly rotating regime \citep{gastine16,long2020}. 

As a conclusion, we observe that variable $\Bio$ convective simulations follow the same general behaviour in term of convective regimes as the two better know end-members.

\section{Discussion}
\label{sec:discussion}

Our objective was to study the influence on thermal convection of Robin boundary condition (characterised by the Biot number) on a rotating spherical shell of fluid heated from below by an isothermal lower interface. This boundary condition has been implemented on the anomalies of temperature and thermal flux at the top interface.

In order to compare with different situations, it appears more convenient to use a more universal Biot number $\Bio^* = Bi/3$ which is based on the fluid height $H = r_e/3$ instead of the radius of our sphere $r_e$.

We used, first, linear stability analysis to determine the properties of this boundary condition through the transition from fixed flux to fixed temperature at the upper boundary (see \ref{sec:lin-stab}). This transition is obtained by varying the Biot number which monitors the linear relationship between the flux and the temperature for a Robin boundary condition at the interface.
We observe a continuous transition in terms of critical Rayleigh number of the system between the two end members.
This transition is significant only at weak rotation for which the two end-member regimes (FF and FT) remain distinct. In the regime of high $\Ek$ (roughly $\Ek>10^{-4}$), a transition occurs approximately for $\Bio^* \in [0.3,30]$. We give a simple model for the evolution of the critical Rayleigh number as a function of $\Bio^*$ during this transition. Outside of this range, the general behaviour of the onset of convection can be described as the traditional FF or FT configuration. These results are in good accordance with previous work of \citet{sparrow_goldstein_jonsson_1964}. In terms of solution structure at the onset of the convection, we offer a schematic phase diagram showing the existence of FF-like, FT-like or intermediate structures depending on $\Ek$ and $\Bio^*$. We have extended our results concerning the FF to FT transition, at least without rotation, to other velocity boundary conditions and other $\Pra$ numbers.

Non-linear simulations have been conducted in a second time (see \ref{sec:nl}) to study vigorous convection.
We have shown that the relevant parameter to compare convective cases subjected to the whole range of boundary conditions authorised by a variable $\Bio^*$ is a Rayleigh number based on the effective mean temperature difference across the shell. We provided a simple model for the evolution of $\Ra_{eff}$ as a function of $\Bio^*$ for non rotating configurations. 

In terms of global diagnostics quantifying convection (for example the Nusselt number or the Péclet number), the transition at a given $\Ra_{eff}$ shows valuable amplitude only for non-rotating and $\Ra_{eff} \simeq \Ra_C$ configurations. 
Indeed, in this case $\Ra_C$ depends significantly on $\Bio^*$, while by construction $\Ra_{eff}$ ignores this effect.
For other values of the parameters, the transition amplitude is considerably reduced. Concerning the organization of the flow, the transition is observable as long as the solution remains not too turbulent (that is $\Ra_{eff} \simeq \Ra_C$). For respectively small and large enough $\Bio^*$ the system behaviour is very similar to the end-member setups. It is only when $\Bio^* \in [3.10^{-2},3.10^{1}]$ that intermediate specific structures can be noticed while global diagnostics remain unchanged quasi identical to those of FF and FT configuration.
Since $\Ra_{eff}$ is not an input parameter, we also studied the transition at fixed $\Ra$ and gave a model for the evolution of $\Nu$ as a function of $\Bio$ (see appendix \ref{sec:tr-ra}).
We observe a noticeable transition only for $\Bio^* \in [3.10^{-2},3.10^{1}]$, this one can be modelled as a simple function of $\Bio$. The transition is in fact characterised by a unique parameter determined by ($\Ra,\Pra,\Ek$).  

Consequently, the range of Biot number for which the results significantly differ from end-member FF or FT boundary conditions is quantified in this paper. Depending on the object of application, computing the Biot number allows one to decide whether it is worth considering the full Robin boundary condition or, otherwise, which end-member model to use. In practice we have shown that Robin boundary conditions can be safely replaced within the framework described in \ref{sec:phymeth} -- by fixed flux boundary condition for $\Bio^*<0.03$ and by fixed temperature for $\Bio^*>30$.

Considering the application to convection in magma ocean planets, we obtained (following the definition given in eq. \ref{defbi} and the references given in \ref{sec:phymeth}) a Biot number of order $10^{10}$ which implies that a fixed temperature at the upper surface is the proper boundary condition to apply.This remains true even if we should bear in mind that a geophysical object like a terrestrial magma ocean evolved with time by cooling and freezing. We must expect that the huge $\Bio^*$ computed here decreases when the ocean evolves and becomes thinner. Indeed $\Bio^*$ is directly proportional to the thickness of the system, inducing a reduction of a factor 100 to 1000 of the Biot number with the crystallization of the ocean.

Nevertheless it is only for very thin layers of high thermal conductivity below a cold atmosphere that FT modelling becomes irrelevant. For a caricatural example, a liquid mercury layer of a hundred of meters below an atmosphere at 300 K sets a Biot number of about 60 which places this system in the intermediate regime. Extreme cases like metric sized liquid layers authorise low Biot number, that is FF boundary conditions. We emphasize that this observation concerns only boundary conditions applied to anomalies and not to the determination of reference profiles which is a distinct work.

However, we also mentioned in the introduction other physical thermal systems which can be modelled using Robin boundary condition. One of them is the superposition of two thermally conductive layers. In that case it is easier to produce configurations for which $\Bio <1$, as thin water ocean (of thickness $H$) below thick ice layer (thickness $H_S$). Except top mechanical boundary condition, our physical model is appropriate to describe this kind of systems (namely internal oceans of icy satellites).
Furthermore, in this situation, the boundary condition depends on the lateral length-scale $\ell$ of the temperature anomaly. For large length-scale, the Biot number is unchanged, but for small length-scales $\ell \ll H_S$ an effective $\Bio_\ell = \Bio H / \ell$ \citep[see][]{guillou95}.
In this context, it may be appropriate to use a Robin boundary condition with a Biot dependent on the length-scale, or at least to use fixed-temperature for the smallest scales of the system.


We have not considered the influence of a symmetric top and bottom Robin boundary conditions or the effects of varying the aspect ratio.
However, our results bear much generality, and a are very likely relevant to most convection setup, as soon as $\Nu$ and $\Ra_{eff}$ are used to characterise the system.

Global and deep oceans of hypothetical exo-planets could constitute another appropriate geophysical subject for such modelling. Robin boundary condition could be chosen to model various relations between a field and its derivative at an interface. We could, for example, contemplate modelling partial crystallization of a two component liquid using this method. Indeed the evolution of the liquid composition $C$ at the solid/liquid interface during the crystallization (due to the fact that the produced solid does not have the same composition as the initial liquid) could be modelled by fixing a chemical flux $j_{C}$ determined by the local composition of the fluid. Such a relation could be made linear by a relation similar to $K\, C + j_{C} = 0$ with $K$ a coefficient similar to the Biot number.

\section*{Acknowledgements}
Computations were performed on the Froggy platform of CIMENT (\url{https://ciment.ujf-grenoble.fr}), supported by the Rh\^one-Alpes region (CPER07\_13 CIRA), OSUG\@2020 LabEx (ANR10 LABX56) and Equip\@Meso (ANR10 EQPX-29-01). 
ISTerre is part of Labex OSUG\@2020 (ANR10 LABX56). 
JV acknowledges funding from the European Research Council (ERC) under the European Union's Horizon 2020 research and innovation program (grant agreement no. 847433).
The open-source numerical codes \textsc{xshells} and \textsc{singe} are available at \url{https://nschaeff.bitbucket.io/xshells/} and \url{https://bitbucket.org/vidalje/singe/downloads/}.
Most figures were produced using matplotlib (\url{http://matplotlib.org/}) or paraview (\url{http://www.paraview.org/}).

\section*{Declaration of interest}
The authors report no conflict of interest.

\appendix

\section{Notations}
\label{sec:anx-nota}

\begin{tabular}{cc}

Parameter & Symbol \\ [3pt]

External radius & $r_e$ \\
  
Aspect ratio & $\eta$ \\
  
Rotative vector & $\mathbf{\Omega}$ \\
  
Gravity acceleration & $\mathbf{g}$ \\
  
Kinematic viscosity & $\nu$ \\
  
Thermal conductivity & $k$ \\
  
Specific heat capacity & $c_p$ \\
  
Density & $\rho$ \\
  
Thermal expansivity & $\alpha$ \\
  
Atmosphere temperature & $T_a$ \\
  
Shell thickness & $L$ \\
  
Thermal gradient accross the shell & $\Delta T$ \\
 
Azimuthal wave number & $m$ \\
 
 Spherical harmonic degree & $l$ \\

\end{tabular} 

\vspace{1cm}

\begin{tabular}{cc}

Field & Symbol \\ [3pt]

Temperature & $T$ \\

Reference conductive temperature & $T_{ref}$ \\

Dimensionless temperature departure from  $T_{ref}$ & $\Theta$ \\

Dimensionless pressure anomaly & $\Pi$ \\

Dimensionless velocity anomaly & $\mathbf{u}$ \\

 \end{tabular} 
 
 \section{Zonal Flows}
 \label{sec:anx-zf}
\begin{figure}
     \centering
     \includegraphics[scale=0.65]{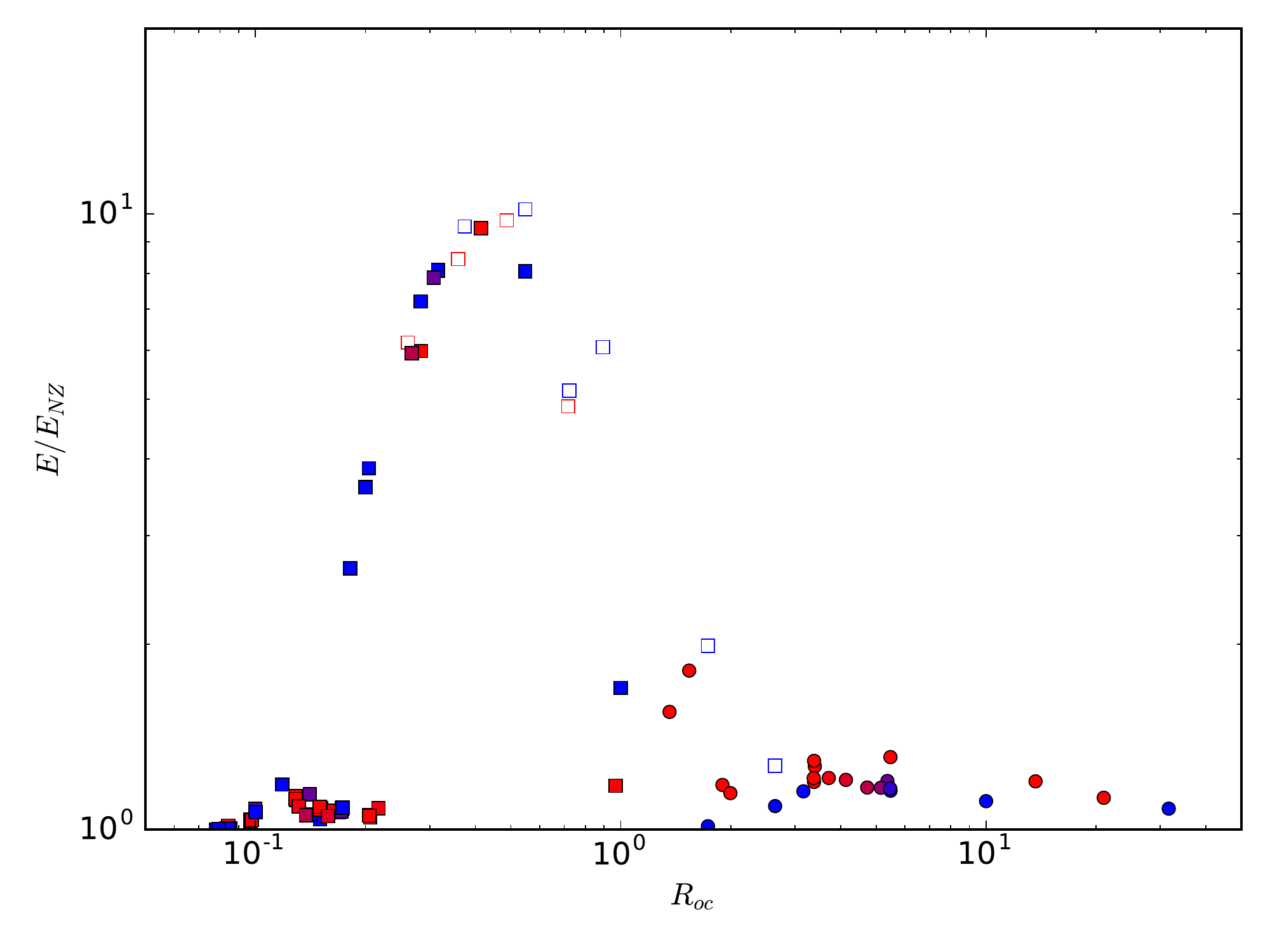}
     \caption{Kinetic energy ratio, $\frac{E}{E_{NZ}}$ as a function of the convective Rossby number $R_{oc}$ for various ($\Ek,\Pra,\Bio$). Legend is similar to fig. \ref{nubi}. Zonal flows are mainly active for small $\Ek$ and $R_{oc}\in [0.1,1]$.}
     \label{znz}
 \end{figure}

 As presented in section \ref{sec:extrnl}, the development of a strong zonal flow is observed for low or intermediate $\Ra$ in rotating simulations. The convective Rossby number $R_{oc}=(\frac{\Ra \ \Ek^2}{\Pra})^{1/2}$ is frequently used as a parameter to study the evolution of zonal flows. $R_{oc}$ estimates roughly the ratio of buoyancy to Coriolis forces \citep{gilman77}.
 Small $R_{oc}$ implies strong influence of rotation while the infinite limit corresponds to a flow not affected by rotation. In order to study this phenomenon we plot on fig. \ref{znz}, the ratio of the total kinetic energy $E$ over the non-zonal component of this energy, $E_{NZ}$, as a function of $R_{oc}$. The larger is the ratio, the stronger is the zonal flow. 
 
 For the two values of $\Ek$, we observed the zonal flow first grows with $R_{oc}$ then decreases.
 Moreover the zonal flow is almost negligible for the weakly rotating configuration (a flat maximum is observed for $R_{oc} \simeq 8$). For $\Ek=10^{-4}$ we observe a maximum for $R_{oc} \simeq 0.5$ at which the zonal flow represent more than $90\%$ of the total kinetic energy of the system. This maximum is compatible with previous results realised with different boundary conditions (see \citet{yadav15} or \citet{christensen2002}). We also observe that $\Pra$ and $\Bio$ do not seem to alter significantly the general behaviour of the zonal flows, only small modification of the maximum being noticed. Finally, fig. \ref{znz} shows that the $\Ek=10^{-4}$ data do not reach very high $R_{oc}$ values ($\sim 3$ at most) which explains that we do not really observe the disappearance of the influence of the rotation on such simulations (see fig. \ref{nu} for example). 
 
 \section{Transition at a given $\Ra$}
  \label{sec:tr-ra}
It was also possible to study the transition observed at constant $\Ra$, when $\Bio$ varies from 0 to infinity. As shown previously this choice is not optimal to compare equivalent configurations but only uses input parameters.

\begin{figure}
\includegraphics[scale=0.65]{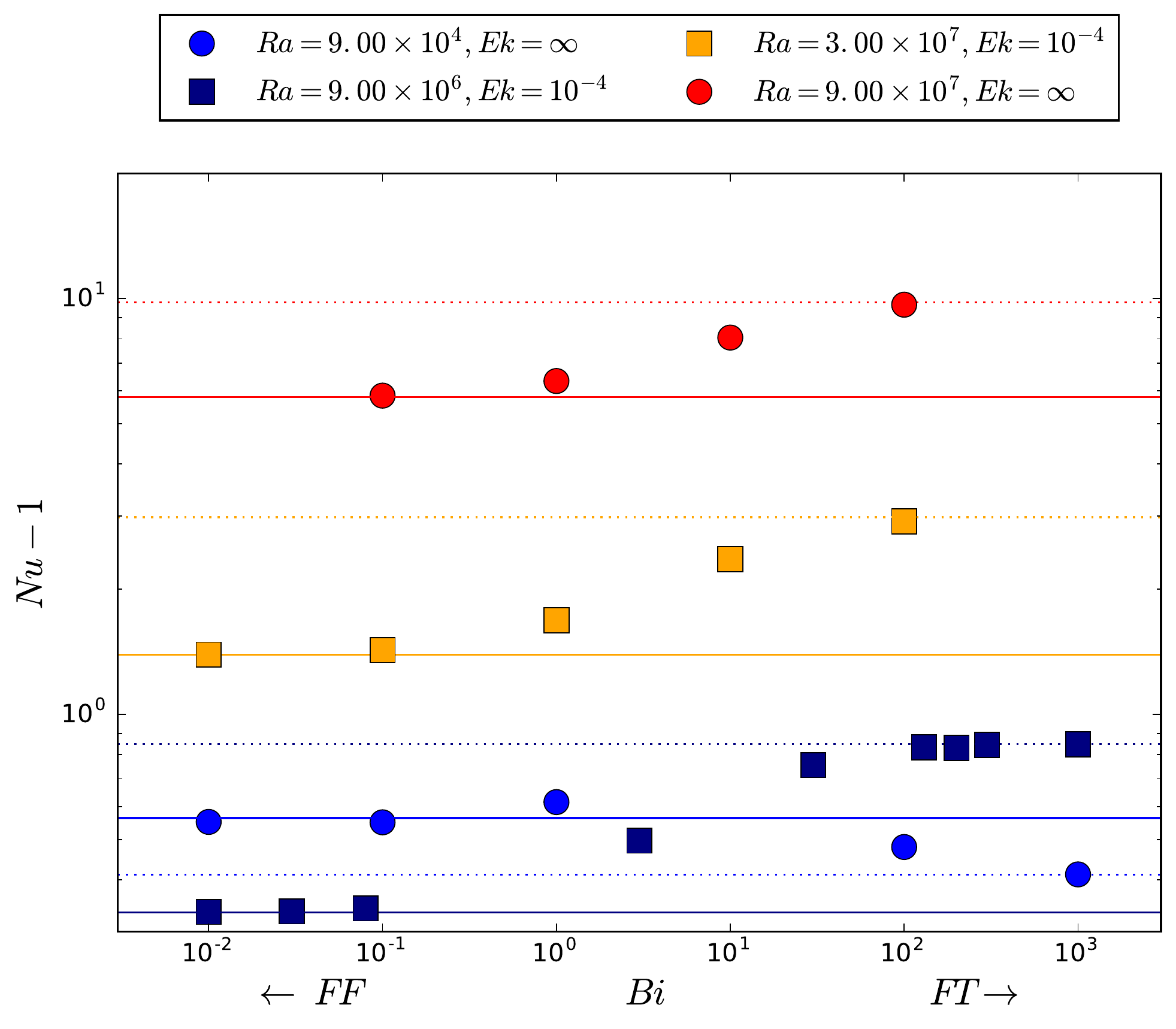}
\caption{Nusselt Number as a function of the Biot Number for various $(\Ra,\Ek,\Pra=3)$ simulations. Each data subset defined by $(\Ra,\Ek)$ is plotted by symbols of the same colour and shape. FF ($\Bio=0$) and FT ($\Bio\rightarrow \infty$) limits are represented respectively by solid and dotted lines of the same colour.}
\label{nubi}
\end{figure}

Fig. \ref{nubi} shows $\Nu-1$ as function of $\Bio$ for various values of ($\Ra,\Ek$). The amplitude of the transition by definition depends on $\Nu(\Bio=0)-\Nu(\Bio=\infty)$ which is a function of ($\Ra,\Ek,\Pra$). For all cases but one, $\Nu$ increases with $\Bio$ -- at equal $\Ra$, FT cases are more efficient in terms of heat transfer. The exception is explained below. Fig. \ref{nubi} shows that in the most common cases the transition from FF to FT is monotonous and continuous. Moreover, the transition occurs mainly between $\Bio=10^{-1}$ and $\Bio=10^{2}$.

\begin{figure}
\includegraphics[scale=0.6]{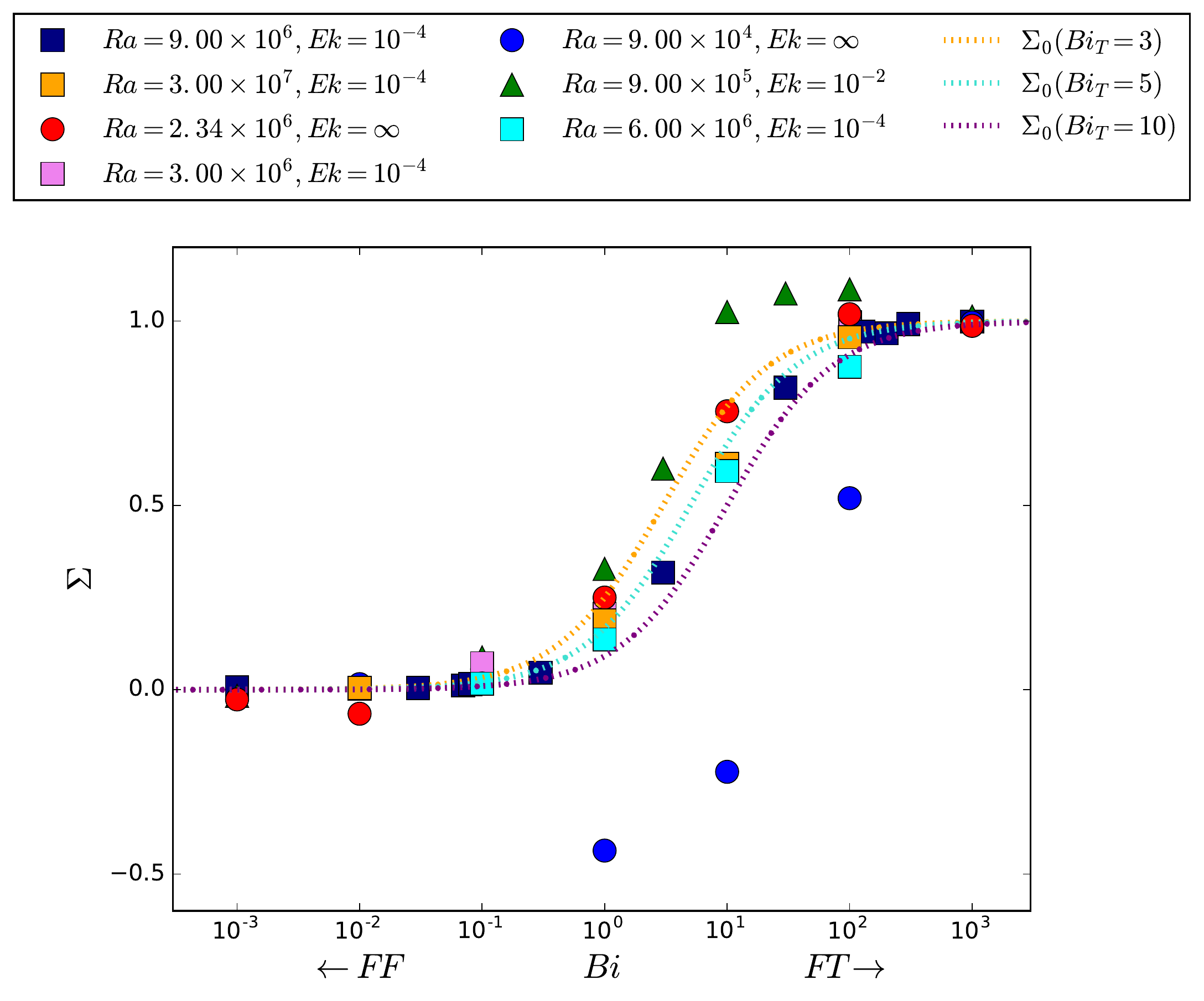}
\caption{$\Sigma$ transition parameter (cf. eq. \ref{eqsig}, computed with $X=\Nu-1$) from FF ($\Bio=0$) to FT ($\Bio=\infty$) as a function of the Biot Number for various $(\Ra,\Ek,\Pra=3)$ simulations. $\Sigma_{0}$ models (cf. eq. \ref{eqsig0}) are plotted for three values of $\Bio_{T}$, $\Bio_T =3,5,10 $}
\label{sigmaet}
\end{figure}

In order to study more specifically the transition itself, we then plot for a given global diagnostics $X(\Ra,\Ek,\Bio)$ (here $X=\Nu-1$) the quantity 
\begin{equation}
\Sigma = \frac{X-X(\Bio=\infty)}{X(\Bio=0)-X(\Bio=\infty)}.
\label{eqsig}
\end{equation}
By construction, $\Sigma$ evolves from 0 to 1 when $\Bio$ goes from 0 to $+ \infty$ and it can be interpreted as a transition parameter. This quantity has been computed for the same subset that was used to produce fig. \ref{nubi} and is plotted on fig. \ref{sigmaet}. First of all, we notice that the large majority of the data subsets associated to a given $(\Ra,\Ek)$ couple follows the same continuous and monotonous transition.
The quantity
\begin{equation}
\Sigma_{0}= \frac{\Bio}{\Bio+\Bio_{T}}
\label{eqsig0}
\end{equation}
approximates correctly the evolution of $\Sigma$, when the central value $\Bio_{T} \approx 3$ to $10$. Note that $\Sigma$ is very similar to the quantity $\gamma$ used to model $\Ra_C=f(\Bio)$ in section \ref{sec:trlin}. We observe a slight  evolution of $\Bio_{T}$ with $\Ra$ and $\Ek$. The only notable exception to these observations is the slightly supercritical, $\Ra = 9\times 10^{4}$ non-rotating case (blue dots), which is anomalous. For this case, the transition is not monotonous and is characterised by a minimum for $\Bio \approx 1$. 
 In fact, this case is one of the few configurations in which the most efficient heat transfer is obtained for intermediate value of $\Bio$. This specific behaviour is linked to the vicinity of the critical Rayleigh number in this simulation as well as to the large relative difference between $\Ra_C$ for FF or FT boundary conditions in this case. As seen in the section \ref{linpart}, $\Ra_C(\text{FF},\Pra,\Ek)<\Ra_C(\text{FT},\Pra,\Ek)$ and $\frac{(\Ra_C\text{FT},\Ek)-\Ra_C(\text{FF},\Ek)}{\Ra_C(\text{FT},\Ek=\infty)-\Ra_C(\text{FF},\Ek=\infty)}$ decreases with $\Ek$. Moreover, $\Nu-1$ is proportional to $\frac{\Ra}{\Ra_C}-1$ near the threshold of convection (\citet{gastine16} and \citet{gillet2006}). Thus, for this case characterised by $(\Ra=9\times 10^{4} \approx \Ra_C,\Ek=\infty,\Pra=3)$ we can reasonably expect that $\Nu(\text{FF})>\Nu(\text{FT})$ which means that increasing $\Bio$ leads to a transition from higher to smaller values. Moreover, since $\Ek=\infty$ the relative amplitude of the transition is large and implies high variations of $\frac{\Ra}{\Ra_C}$ with $\Bio$ since $\Ra$ remains constant. This produces the non-monotonous variation of $\Nu$ with $\Bio$.

We also consider other global diagnostics: $\langle Et \rangle$ which is the root mean square value of the temperature anomaly and $P_{conv}$ which is the root mean square of $u\Theta$ and found similar behaviour as for $\Nu$ with respect to $\Bio$.

  \section{3D representations}
\label{sec:anx-3D}

\begin{figure}

\begin{tabular}{cc}

i)\ \ \includegraphics[scale=0.25]{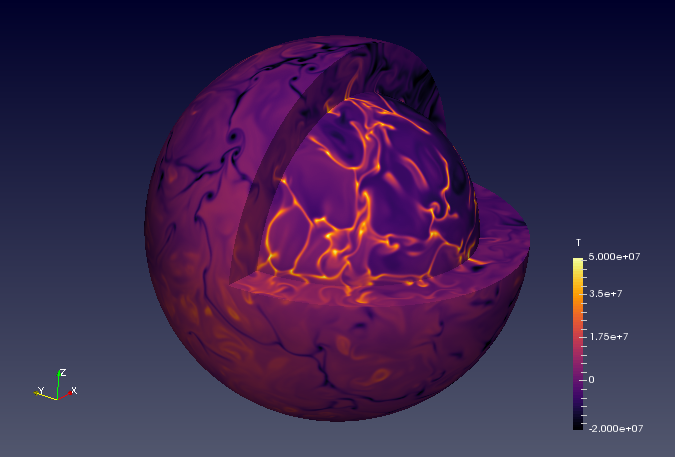}& ii) \includegraphics[scale=0.25]{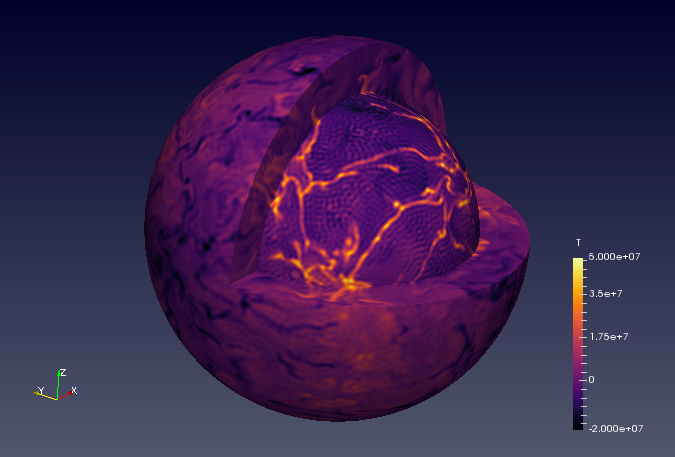}\\ 

iii) \includegraphics[scale=0.25]{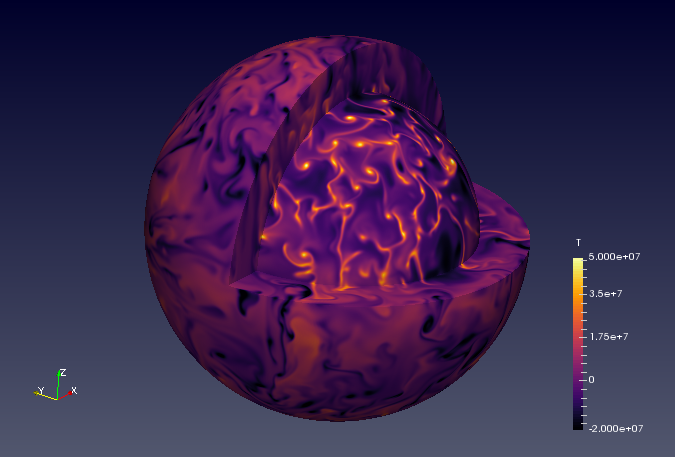}& iv) \includegraphics[scale=0.25]{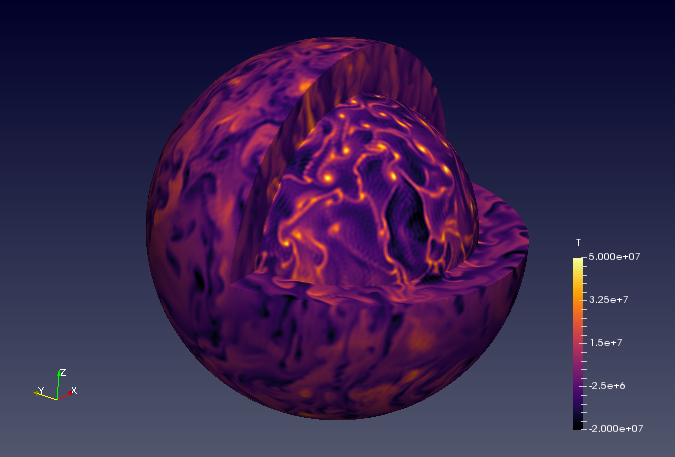}\\ 

v)\ \ \includegraphics[scale=0.25]{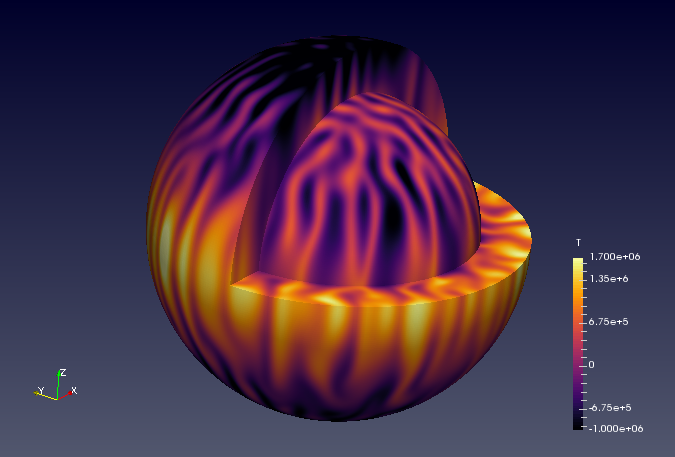}& vi) \includegraphics[scale=0.25]{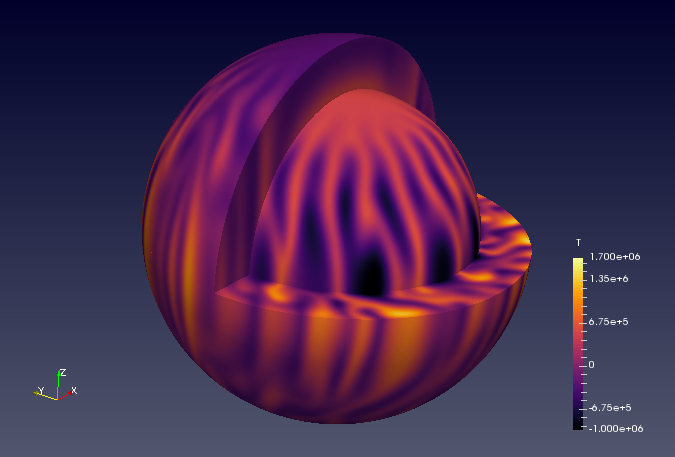}\\ 

\end{tabular} 
\caption{3D representation of the temperature anomaly (l=0 subtracted) inside a shell of radial limits 0.7 and 0.95 with $\Pra=3$. Rotational axis is z. Left column FF, right one FT. For each line ($\Ra_{eff},\Ek$) is kept approximately constant. First line ($3\times10^{8},\infty$). Second line ($3\times10^{8},10^{-4}$). Third line ($10^{7},10^{-4}$).}
\label{fig:3D}
\end{figure}

Fig. \ref{fig:3D} shows 3D maps of temperature anomalies (l=0 component subtracted) for sub-shells of our system -- boundary layers excluded -- for $\Pra=3$.
FF and FT configurations are compared for quasi equal $Ra_{eff},\Ek$. For cases with $\Ra_{eff}=3\ 10^8$, the effect of boundary condition is hardly visible. Conversely, cases with lower $\Ra_{eff}$ with a strong effect of rotation show differences in the location of the anomalies with respect to the tangent cylinder (see fig. \ref{fig:U-s-Pr} I).

\bibliographystyle{jfm}
\bibliography{biblio.bib, jfm-instructions.bib}

\end{document}